\documentclass[aoas,preprint]{imsart}

\RequirePackage{amsthm,amsmath,amsfonts,amssymb}
\RequirePackage[authoryear]{natbib}
\RequirePackage[colorlinks,citecolor=blue,urlcolor=blue]{hyperref}
\RequirePackage{graphicx}

\startlocaldefs
\theoremstyle{plain}

\theoremstyle{remark}

\endlocaldefs

\usepackage{xfp}
\newcommand\SupplementaryMaterials{%
  \xdef\presupfigures{\arabic{figure}}
  \xdef\presupsections{\arabic{section}}
  \renewcommand\thefigure{D\fpeval{\arabic{figure}-\presupfigures}}
}

\begin{document}
\begin{changemargin}{-1cm}{-9mm}{-5mm}
\setlength{\textheight}{8.6in}
\setlength{\paperheight}{279mm}
\setlength{\paperwidth}{216mm}

\begin{frontmatter}
\title{Spatiotemporal Modeling of Nursery Habitat Using Bayesian Inference: Environmental Drivers of Juvenile Blue Crab Abundance}
\runtitle{Modeling Blue Crab Abundance}

\begin{aug}
\author[A,C]{\fnms{A. Challen} \snm{Hyman}\ead[label=e1]{achyman@vims.edu}},
\author[A,B,C]{\fnms{Grace S.} \snm{Chiu}\ead[label=e2]{gschiu@vims.edu}},
\author[A]{\fnms{Mary C.} \snm{Fabrizio}}
\and
\author[A]{\fnms{Romuald N.} \snm{Lipcius}}
\address[A]{Virginia Institute of Marine Science, William \& Mary, Gloucester Point, VA 23062, USA}

\address[B]{Australian National University; Virginia Commonwealth University; University of Washington; University of Waterloo}

\address[C]{Corresponding authors: \printead{e1}; \printead{e2}}
\end{aug}

\begin{abstract}
Nursery grounds are favorable for growth and survival of juvenile fish and crustaceans through abundant food resources and refugia, and enhance secondary production of populations. While small-scale studies remain important tools to assess nursery value of habitats, targeted applications that unify survey data over large spatiotemporal scales are vital to generalize inference of nursery function, identify highly productive regions, and inform management strategies. Using 21 years of GIS and spatiotemporally indexed field survey data on potential nursery habitats, we constructed five Bayesian models with varying spatiotemporal dependence structures to infer nursery habitat value for juveniles of the blue crab C. sapidus within three tributaries in lower Chesapeake Bay. Out-of-sample predictions of juvenile counts from a fully nonseparable spatiotemporal model outperformed predictions from simpler models. Salt marsh surface area, turbidity, and their interaction showed the strongest associations (and positively) with abundance. Relative seagrass area, previously emphasized as the most valuable nursery in small spatial-scale studies, was not associated with abundance. Hence, we argue that salt marshes should be considered a key nursery habitat for blue crabs, even amidst extensive seagrass beds. Moreover, identification of nurseries should be based on investigations at broad spatiotemporal scales incorporating multiple potential nursery habitats, and on rigorously addressing spatiotemporal dependence.
\end{abstract}

\begin{keyword}
\kwd{blue crab}
\kwd{nursery habitat}
\kwd{\textit{Callinectes sapidus}}
\kwd{Bayesian hierarchical model}
\kwd{spatiotemporal model}
\kwd{salt marsh}
\kwd{seagrass}
\kwd{ecosystem based fishery management}
\end{keyword}

\end{frontmatter}

\section{Introduction}

A key element of ecosystem-based fishery management (EBFM) is the incorporation of habitat (e.g., EFH, ``Essential Fish Habitat'') into management, conservation and restoration decisions \citep{MSA2007}. However, quantitative assessments of the production value of habitats have only recently been attempted \citep{vasconcelos2014patterns, Seitz2014, wong2016, brown2019, camp2020}; see  \citet{lipcius2019modeling} for a review. In particular, nursery habitats can enhance growth and survival of juvenile fish and crustaceans in diverse marine and estuarine ecosystems \citep{beck2001identification, heck2003critical, minello2003salt, nagelkerken2015seascape, litvin2018makes, yoklavich2010marine, peters2018habitat} through the provision of food resources and refugia. Hence, linking nursery habitat quantity and quality to population dynamics and EBFM of exploited species has been emphasized \citep{Seitz2014, vasconcelos2014patterns, wong2016, brown2019, lipcius2019modeling, camp2020}. 

Unfortunately, quantification of habitat value has been uncommon due to the considerable logistical difficulties associated with field experiments \citep{beck2001identification}. Until recently, comparison of potential nurseries relied primarily on examination of specific habitat types (e.g., sea grass, oyster reef, marsh) as single units disconnected from adjacent habitats \citep{nagelkerken2015seascape}. However, estuaries are complex habitat mosaics that include physical, biotic, and chemical components interacting at multiple spatial and temporal scales \citep{olson2019nearshore}, and as such, these connections must be considered. Operational definitions of nurseries must be expanded to consider multiple structured and unstructured habitat types, as well as environmental characteristics within a region \citep{nagelkerken2015seascape}. Furthermore, inference on nursery habitat value is complicated in that habitat preferences of many marine and estuarine species change with ontogeny, such that early-life stages frequently inhabit different habitats than older juveniles or adults \citep{jones2010connectivity, nakamura2012variability, epifanio2019early}. Quantitative assessments of nursery function and fisheries production must therefore move beyond comparisons between specific habitat types \citep{nagelkerken2015seascape, sheaves2015true, litvin2018makes} and be considered within the context of ontogeny \citep{lipcius2007post}, especially for organisms with complex life cycles \citep{lipcius2007post, epifanio2007biology, epifanio2019early}.

While ecological studies often quantify nursery function at fine temporal and spatial scales, few are conducted at the scales relevant to the population \citep{turner2008ecosystem}. Small-scale studies on the importance of structured habitats as nurseries may not scale up to the population level. For example, high local juvenile density or survival in small-scale studies \citep{beck2001identification} may not translate to high secondary production in a population if per-unit-area productivity of a potential nursery habitat is negated by the small area of a habitat in the seascape \citep{dahlgren2006marine, nagelkerken2015seascape}. For robust evaluation of nursery habitats at sub-population or population scales, small-scale field studies should be complemented with analyses of large-scale field data, especially when informing decision-making within the context of EBFM. 

The blue crab \textit{Callinectes sapidus}, which supports valuable fisheries along the Western Atlantic and Gulf of Mexico coasts \citep{NOAALANDINGS2019}, is a model organism for quantifying value of structured habitats under spatially and temporally varying environmental characteristics. Like many exploited marine species, the blue crab utilizes a range of nursery habitats and exhibits ontogenetic shifts in habitat utilization \citep{orth1987utilization, hines2007ecology, lipcius2007post,Seitz2014, epifanio2019early}. Postlarvae settle in structured habitats, such as seagrass, where they metamorphose to the first juvenile instar \citep{metcalf1992relationship} and either remain or exhibit density-dependent secondary dispersal to alternative structured habitats \citep{etherington2000large,etherington2003partitioning, johnston2012exotic}. After reaching 20-25 mm carapace width (CW), they emigrate to unstructured soft-bottom habitats \citep{lipcius2005density, seitz2005food}, but also continue to use structured habitats for foraging, molting, and mating \citep{hines2007ecology, lipcius2007post}. For the blue crab, \citet{hines2007ecology} and \citet{lipcius2007post} reviewed the extensive evidence for the value of specific nursery habitats, such as seagrass, using the definition of nursery habitat as areas with elevated per-unit-area density, survival and growth. 

Two aspects of the blue crab's life history are particularly useful in quantifying value of nursery habitats. First, size-specific habitat use and dispersal patterns of the blue crab through ontogeny are well understood \citep{lipcius2007post,hines2007ecology}. Second, male and immature female blue crabs larger than 20 mm carapace width (CW) exhibit high site fidelity and low emigration rates during summer and fall at spatial scales less than a few kilometers \citep{wrona2004determining, davis2004comparing, hines2008release, johnson2010population}. Hence, abundance of juvenile blue crabs larger than 20 mm CW can be used to identify areas of high productivity, and facilitate quantitative comparisons of the relative contribution of multiple nursery habitats in the seascape to the population.

Here, we exploited the differential habitat utilization of pre- and post-dispersal juvenile blue crabs to infer relative nursery value of various habitats associated with specific environmental characteristics. We constructed statistical spatiotemporal models to examine geographic heterogeneity in post-dispersal juvenile blue crab abundance and to infer variation in nursery habitat value within and across estuaries in lower Chesapeake Bay, Virginia. Specifically, we used temporal extensions of conditional autoregressive (CAR) spatial models to assess the effects of environmental factors on abundance of juvenile blue crabs at the tributary and regional scales simultaneously. Using local abundance of post-dispersal sized (20-40 mm CW) juvenile blue crabs as an indicator for local production, our objectives were to 1) evaluate relationships between nursery habitat distribution and local productivity at regional scales ($\geq$ 100 km\textsuperscript{2}), and 2) identify geographic areas with consistently high abundance and productivity.

\section{Study design}
\subsection{Study Area}
The three large tributaries analyzed in this study–the James, York, and Rappahannock Rivers–discharge into the lower portion of western Chesapeake Bay and serve as nursery, foraging and spawning habitats for many marine and estuarine species (Fig. \ref{fig:Map}). The tributaries are partially mixed, coastal plain subestuaries with depths generally between 5 to 10 m along the axes, but with deeper portions ($>$20 m) near the mouths \citep{smock2005atlantic}. Each tributary contains a range of seagrass and salt marsh configurations. Seagrasses, primarily eelgrass \textit{Zostera marina} and widgeon grass \textit{Ruppia maritima}, vary from large, continuous meadows to areas with few small patches of variable shoot densities \citep{hovel2002effects}. Salt marshes, dominated by smooth cordgrass \textit{Spartina alterniflora}, span extensive sections of the shorelines of each tributary, although areal coverage of marsh patches varies spatially among and within individual tributaries. 

\begin{figure}[ht]
\center{\includegraphics[width=\textwidth]{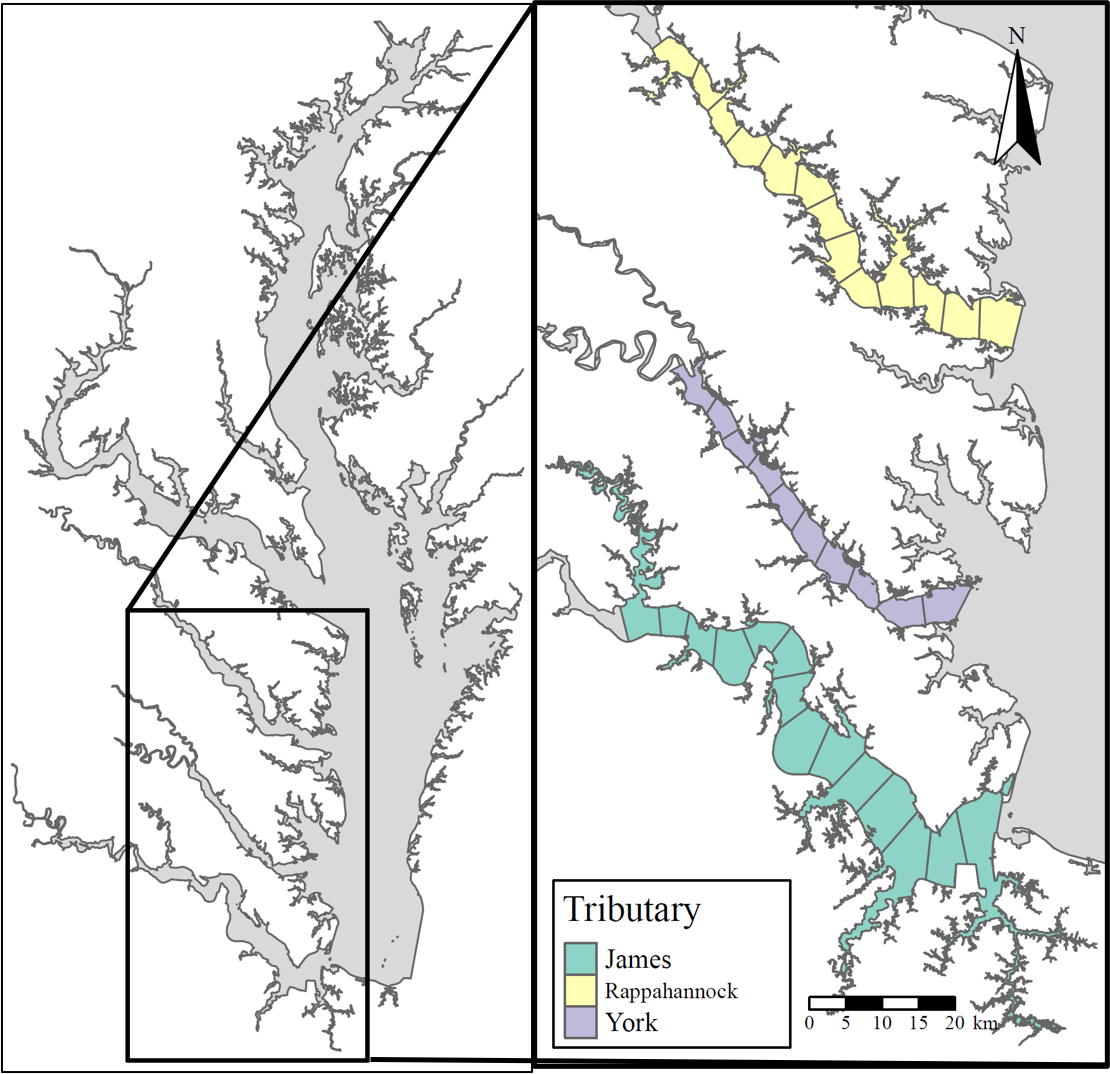}}
\caption{Map of Rappahannock, James, and York rivers with tributary sections (areal units) superimposed. See Section \ref{sec:sampling} and Appendix \ref{appendix:areal units} for the definition of areal units within tributaries.}
\label{fig:Map}
\end{figure}

\subsection{Predictors of Abundance and Productivity}
Seven environmental variables (herein, predictors) were initially considered as potential determinants of local productivity for juvenile blue crabs, and are described below. Additional details on variable definition and associated regression coefficients are in Section \ref{sec:sampling}, Table \ref{table:description}, and Appendix \ref{appendix:parameters}. We did not include salinity as a predictor due to substantial collinearity with turbidity and location along the river axis.

\begin{table}[htbp]
\caption{Descriptions of predictors used in all initial models} \label{table:description}
\begin{center}
\begin{tabular}{l l p{7cm}}
  \hline
 Predictor & Regression & Description\\ 
 & Coefficient & \\
  \hline
  $\qquad$ --- & $\beta_0$ & (Intercept of model) \\ 
  Tow distance (log) & $O$ & Offset term relating juvenile blue crab abundance to surface area river sections \\
  Turbidity & $\beta_{\text{Turbidity}}$ & Mean water cloudiness measured as the negative value of the Secchi disk depth (m) for the $k$th river section in the $t$th year\\ 
  Seagrass (relative area) &  $\beta_{\text{Seagrass}}$ & $\frac{(\text{SAV area})_{kt}}{(\text{Section area})_{k}}=$ Total area of SAV in section $k$ at time $t$ divided by the area of section $k$ \\ 
  Marsh (relative area) &  $\beta_{\text{Marsh}}$ & $\frac{(\text{Marsh area})_{kt}}{(\text{Section area})_{k}}=$ Total area of salt marsh in section $k$ at time $t$ divided by the area of section $k$ \\ 
  Marsh $\times$ Turbidity &  $\beta_{\text{Marsh} \times \text{Turbidity}}$ & Interaction term between marsh relative area and turbidity.\\ 
  Predator abundance (log-count) &  $\beta_{\text{Predator}}$ & Log-transformed counts, in section $k$ at time $t$, of predator abundance between 100 and 300 mm in total length (fish) or CW (adult blue crabs) from the 10 most common predators of small juvenile blue crabs (see Table \ref{table:Species}) \\ 
  Management (post 2008) &  $\beta_{\text{Management}}$ & Effect of Chesapeake Bay blue crab management changes enacted in 2008 \\ 
  Rappahannock &  $\beta_{\text{Rappahannock}}$ & Tributary-specific effect of the Rappahannock River relative to the James River (baseline)\\ 
  York & $\beta_{\text{York}}$ & Tributary-specific effect of the York River relative to the James River (baseline)  \\ 
   \hline
\end{tabular}
\end{center}
\end{table}

\subsubsection*{Seagrass}
Historically, emphasis was placed on seagrass meadows as the preferred nursery for small (i.e., $<$30 mm CW) juvenile blue crabs \citep{orth1987utilization, perkins1996nursery, hovel2002effects, ralph2013broad} due to the high densities of juvenile crabs and settlement of postlarvae \citep{olmi1990variation, welch1997effects, van2003substrate} in seagrass meadows over alternative structured and unstructured substrates \citep{orth1987utilization, lipcius2005density}. Effects of seagrass area are likely influenced by the spatial extent of the river section. Thus, we defined a relative seagrass area metric by dividing the area of seagrass cover within each river section in each year by the area of that section to yield a relative seagrass area metric (i.e., percent area covered, PAC). Hereafter, we refer to the spatiotemporal unit representing a given river section in each year as a section-year. 

\subsubsection*{Marsh}
Salt marshes may serve as nursery habitat for juvenile blue crabs, particularly in locations where seagrass is absent or declining \citep{jivoff2003evaluating,bishop2010blue,johnson2010population}. In the Gulf of Mexico, juvenile blue crab abundance is high in both seagrass and salt marsh habitats \citep{thomas1990abundance,rozas1998nekton, heck2001pre}. In tethering experiments, survival of juveniles was comparable between the two habitats, both of which had higher survival than in unstructured habitat \citep{shakeri2020blue}. Similar to seagrass, we defined a relative marsh area metric for each section-year.

\subsubsection*{Turbidity}
Strong turbidity gradients exist in each tributary \citep{nichols1973development,kuo1978modeling,lin2003model,filippino2017influence}. Dissolved and particulate suspended solids are imported from surrounding watersheds to tributaries via terrestrial runoff. In contrast, seawater from estuarine mouths is relatively clear. Divergence in turbidity is apparent when comparing marine (i.e., polyhaline) to mesohaline and oligohaline, highly turbid upriver areas, where water clarity is frequently driven by allochthonous inputs and sediments from the surrounding watershed. The estuarine turbidity maximum (ETM), a region of elevated suspended solid concentrations and reduced light availability, occurs near the limit of salt intrusion in each tributary, where turbidity peaks \citep{sanford2001reconsidering}.

Turbidity may provide juvenile blue crabs with protection from visual predators \citep{cyrus1987influence,marley2020mangrove} and from cannibalism by larger congeners \citep{o1976apparent} through a reduction in light intensity. Upriver unstructured habitat is turbid, whereas similar habitat downriver has lower turbidity, such that upriver unstructured habitat can also serve as an effective nursery \citep{lipcius2005density,seitz2005food}. Hence, mean turbidity per section-year was included as a continuous covariate.

\subsubsection*{Marsh-Turbidity Interaction}
Whereas seagrass meadows do not occur in high-turbidity areas due to light requirements, extensive salt marshes are present in both high- and low-turbidity regions of the tributaries. As such, turbidity may modify the effectiveness of structured salt marsh habitat as nursery for juvenile crabs by decreasing predatory foraging efficiency through both low visibility (turbidity) and structural impediments (marsh grass). For this reason, the interaction between marsh area and turbidity was included in the analysis. We recognize that there may be confounding variables with turbidity, such as location along the upriver-downriver gradient and resources such as prey availability. Hence, our interpretations will be limited to a association between crab abundance and turbidity.

\subsubsection*{Predation}
Although physical refuges can reduce predator foraging success, predator density may also determine survival \citep{Mintz1994}. For example, high abundances of juvenile blue crabs in low salinity regions have been linked to low predator abundance in those regions \citep{posey2005importance}. Predators for each section-year were determined from the literature and abundances of the 10 most important predators of small juvenile blue crabs, including larger conspecifics (Table \ref{table:Species}), were obtained from the Virginia Institute of Marine Science Juvenile Fish and Blue Crab Trawl Survey (hereafter VIMS Trawl Survey) \citep{Trawl2019}. Predator densities were estimated for individuals between 100 and 300 mm in total fish length (or CW for blue crabs), with the lower bound defining the smallest size able to capture and consume small juvenile blue crabs \citep{scharf2000predator}, and the upper bound representing animals which would be expected to consume smaller juvenile blue crabs.

\begin{table}[ht]
\caption{List of predator species considered in predation abundance variable}
\begin{center}
\renewcommand*{\arraystretch}{1.2}
\begin{tabular}{p{0.2\linewidth} p{0.28\linewidth} p{0.5\linewidth}}
  \hline
 Common name & Species name &  Source \\ 
  \hline
  Blue crab (adult)& \textit{Callinectes sapidus}  & \citet{hines2007ecology,bromilow2017mechanisms} \\ 
  Striped Bass &\textit{Morone saxatilis}  & \citet{hines2007ecology, lipcius2007post,bromilow2017juvenile}  \\ 
  Red Drum &\textit{Sciaenops ocellatus} &  \citet{hines2007ecology,guillory2001review,bromilow2017juvenile} \\ 
  Silver Perch &\textit{Bairdiella chrysoura} & \citet{guillory2001review} \\
  Weakfish &\textit{Cynoscion regalis} & \citet{bromilow2017juvenile} \\ 
  Atlantic Croaker &\textit{Micropogonias undulates} & \citet{guillory2001review,bromilow2017juvenile} \\ 
  Northern Puffer &\textit{Sphoeroides maculatus} & \citet{bromilow2017mechanisms} \\ 
  Striped Burrfish &\textit{Chilomycterus schoepfi} & \citet{bromilow2017mechanisms} \\ 
  Blue Catfish &\textit{Ictalurus furcatus} & \citet{schmitt2019modeling} \\
  Oyster Toadfish &\textit{Opsanus tau} & \citet{bromilow2017mechanisms}\\
   \hline
\end{tabular}
\end{center}
\label{table:Species}
\end{table}

\subsubsection*{Tributary}
The three tributaries in our study vary in geography, morphology, and hydrology. Average discharge is relatively high in the James River (194 m$^3$ s$^{-1}$) and lower in the Rappahannock and York Rivers (47 m m$^3$ s$^{-1}$  and 31 m$^3$ s$^{-1}$, respectively) \citep{cronin1971volumetric}. Additionally, the three tributaries are positioned along a latitudinal gradient in Chesapeake Bay, with the Rappahannock River being northernmost, the James River southernmost and closest to the Bay mouth, and the York River intermediate. Finally, these tributaries vary substantially in surface area \citep{cronin1971volumetric}. The James River is the largest at 513.0 km\textsuperscript{2}, the York River is the smallest  at 162.5 km\textsuperscript{2}, and the Rappahannock River is intermediate at 307.5 km\textsuperscript{2} \citep{smock2005atlantic}. Variation in these physical characteristics may affect blue crab abundance and thus, tributary was considered as a predictor in the model.

\subsubsection*{Management}
Early in the 1990s the blue crab spawning stock in Chesapeake Bay declined by 80\% \citep{lipcius2002concurrent}, and average annual female abundance dropped 50\% from 172 million crabs in 1989-1993 to 86 million crabs in 1994-2007 \citep{MDNR2019}. As a consequence, larval abundance and postlarval recruitment were lower by approximately 1 order of magnitude \citep{lipcius2002concurrent}. The sharp decline resulted in a range of management and recovery actions implemented from 2001 through 2008, including establishment of an extensive spawning sanctuary that encompassed about 75\% of the spawning grounds in Chesapeake Bay \citep{lipcius2001deepwater, lipcius2003spatial,lambert2006assessing}. Most notably, severe fishery reductions were implemented in 2008 by the three management agencies, which included the Virginia Marine Resources Commission, Potomac River Fisheries Commission, and Maryland Department of Natural Resources (MDNR), leading to a 34\% reduction in female landings across Maryland and Virginia \citep{MDNR2019} and triggering population recovery in subsequent years. Since 2008, annual female abundance rebounded to pre-1994 levels, and stabilized at an average of 161 million crabs during 2008-2019 \citep{MDNR2019}. We included management status (before and after 2008, with 2009 being the first recruitment period after management change) as a categorical predictor to capture potential effects on juvenile blue crab abundance due to increases in female blue crab abundance in response to regulatory changes that were implemented in 2008. However, we also realize that the effects of management may be interactive with those of other factors (e.g., management may increase abundance in marsh habitats but not in unvegetated areas), and thus we interpret the results for the additive effect of management with caution.

\subsection{Sampling and Data Processing}\label{sec:sampling}
Juvenile blue crab and predator abundance data were obtained from the fisheries-independent VIMS Trawl Survey \citep{Trawl2019}. Beginning in March 1996 and continuing to the present, stratified‐random and fixed-site sampling has been conducted monthly in the James, York, and Rappahannock Rivers using consistent gear, research vessel, and methodology. Secchi disk measurements, a proxy for turbidity, are collected immediately following each trawl tow. This sampling design provided a monthly time series of juvenile blue crab and predator catch data as well as water quality data (temperature, turbidity) in each tributary. The maximum size of predators (300 mm fish total length or crab carapace width) represent the sizes that are reliably captured by the VIMS Trawl Survey \citep{Trawl2019}.

GIS data on submersed aquatic vegetation (SAV) and salt marsh distributions were used as explanatory habitat variables. SAV polygons digitized from annual aerial photographs were obtained from the \href{http://www.vims.edu/bio/sav}{VIMS SAV program}, while polygons of salt marsh distributions were obtained from the \href{https://www.vims.edu/ccrm/research/inventory/index.php}{Shoreline and Tidal Marsh Inventory dataset} from the VIMS Center for Coastal Resource Management. 

The spatiotemporally varying samples from the VIMS Trawl Survey were aggregated to annually indexed areal units for the period 1996 to 2017. We limited the months considered for each year to April--December to avoid bias in crab distributions associated with winter dormancy \citep{hines2007ecology}. First, each tributary was divided along its axis into sections approximately five km in length resulting in $K= K_1+K_2+K_3=37$ total areal units, or sections ($K_1=14$ for James, $K_2=13$ for Rappahannock, and $K_3=10$ for York), which excluded one polygon at the mouth of the James representing the first five km because no samples were collected in this region by the trawl survey (Fig. \ref{fig:Map}). Areal unit definitions are discussed in Appendix \ref{appendix:areal units}. For each $k$th areal unit within each $t$th year (i.e., $(k,t)$th section-year), blue crab catch and tow distance (m) information were summed to derive values of total abundance and total tow distance. Secchi disk depth and log\textsubscript{e}-transformed predator abundance values were averaged within each $(k,t)$. Finally, marsh and seagrass area within each section-year divided by the total area of each section were used as a relative habitat area metric for each structurally complex habitat. The aggregated trawl data resulted in 814 section-year observations. All but one of the 814 section-years contained trawl tows, and the exception was from 2017. This aggregation resulted in values of relevant response and predictor variables for each river section $k$ in each year $t$.

\section{Model Development and Specification}\label{meth:development}
The spatiotemporal structure of the data in this study required complex modeling because sampling sites did not represent independent replicates. The study region covered three tributaries, each comprised of a set of $k = 1, . . . , K_g$ non-overlapping areal units (sections), $g=1,2,3$, and data were recorded for each section for $t = 1, . . . , T$ consecutive time periods ($T=21$ years over 1996--2016, due to the 2017 data being withheld for out-of-sample cross validation; see Section \ref{sec:validate}). A multilevel (hierarchical) spatiotemporal Bayesian model framework for discrete responses (count data) was used to evaluate the effects of predictors while simultaneously accounting for spatiotemporal dependence. We used temporal extensions of conditional autoregressive (CAR) models \citep{waller1997hierarchical} to examine spatiotemporal patterns in the abundance of juvenile blue crabs among potential nursery areas. To determine the necessary model complexity to capture spatial and temporal patterns in juvenile blue crab abundance, we constructed five model variants with various spatiotemporal dependence structures. Specifically, we compared models that i) ignored spatial and temporal autocorrelation, ii) considered exclusively spatial autocorrelation, iii) considered separable (i.e., non-interacting) spatial and temporal autocorrelation (split into (3a) and (3b)) and iv) considered fully non-separable (i.e., interacting) spatiotemporal autocorrelation.

\subsection{Model 1}\label{meth:model 1}
The simplest model considered in this study was a Poisson generalized linear mixed-effects model with a random effect for all river sections $k=1,...,K$:

\begin{flalign}
Y_{kt} | \mu_{kt} &\sim \text{Pois}(\mu_{kt}) \\
\ln(\mu_{kt}) &= \sum_{i=0}^{p} x_{kti} \beta_i + O_{kt} + \theta_k \medspace\qquad\qquad\nonumber\\
&\theta_k|\sigma_\theta^{2} \sim N(0,\sigma_\theta^2)\medspace\nonumber\\
& \qquad \beta_i \sim N(0,100) \nonumber\\
& \qquad \sigma_\theta^2 \sim \text{inverse-Gamma}(1,1) \nonumber
\end{flalign}

The response data, juvenile crab counts, are denoted by $Y_{kt}$, for the $k$th section in year $t$. Tow distances, known offsets that have been log-transformed, are denoted by $O_{kt}$. An offset variable is one that is treated like a regression covariate whose slope parameter is fixed at 1. Offset variables are most often used to scale the modeling of the mean structure when the response variable is expected to be proportional to the offset term. A vector of predictors (see Table \ref{table:description}), $x_{kt} = (1, x_{kt1}, . . . , x_{ktp})$ is denoted for each $(k,t)$, and includes $x_{kt0}=1$ which corresponds to the intercept term. Model 1 included an independent and identically distributed (i.i.d.) random effect, $\theta_k$. This parameter was normally distributed and accounted for section-specific variation only and did not consider spatial autocorrelation among neighboring sections or temporal autocorrelation within a given section through time. All fixed-effect regression coefficients were given a normal prior distribution with mean 0 and variance 100. The random-effect variance $\sigma^2_\theta$ was given an $\text{inverse-Gamma} (1,1)$ hyperprior, which is reasonably diffuse to reflect the lack of information about the parameter.

\subsection{Model 2}\label{meth:model 2}
This model considered the effects of spatial autocorrelation among neighboring river sections through the substitution of i.i.d. $\theta_k$ with conditionally autoregressive (CAR) $\Phi_k$ \citep{ver2018spatial}:

\begin{flalign}
Y_{kt} | \mu_{kt} &\sim \text{Pois}(\mu_{kt}) \\
\ln(\mu_{kt}) &= \sum_{i=0}^{p} x_{kti} \beta_i + O_{kt} + \Phi_{k} \nonumber\\
&\Phi|\Sigma \sim \text{MVN}(0,\Sigma)\medspace\nonumber\\
& \qquad \Sigma = \sigma_\Phi^2 (D - \lambda W)^{-1} \nonumber\\
& \qquad \beta_i \sim N(0,100) \nonumber\\
& \qquad \lambda \sim \text{U}(0, 1)\nonumber\\
& \qquad \sigma^2_\Phi \sim \text{inverse-Gamma}(1,1) \nonumber
\end{flalign}

where the joint probability distribution of $\Phi = (\Phi_1, . . . , \Phi_K )$ is specified as a multivariate normal distribution with a mean vector of 0s and variance-covariance matrix $\Sigma$. The $\Sigma$ matrix describes spatial correlation based on the neighborhood structure specified by a $K \times K$ adjacency matrix, $W$, and an autocorrelation parameter $\lambda$, which controls the degree of spatial autocorrelation among neighboring sections across the entire region of study. We employed a binary weighting scheme for $W$ where $w_{k,k'}=0$ for all $(k,k')$ unless areal units $k\ne k'$ share a common border. The influence of neighboring sections on a given section was standardized by subtracting $\lambda W$ from $D$, a diagonal matrix where $D_{k,k}$ is the number of neighbors for section $k$. This specification effectively scaled spatial dependence by the number of neighbors for each section while avoiding model unidentifiability of the intrinsic CAR (ICAR) structure \citep[e.g.,][]{chiu2013spatial}. The parameter $\lambda$ was constrained between 0 and 1 (hence, non-negative) through a uniform prior. This spatial dependence structure was assumed to be homoscedastic through the variance parameter $\sigma^2_\Phi$, again with an $\text{inverse-Gamma} (1,1)$ hyperprior. The regression coefficients were given the same prior distributions as before.

\subsection{Model 3}\label{meth:model 3}
Models 3a and 3b considered separable spatial and temporal dependence \citep{waller1997hierarchical} by expanding on Model 2 through the addition of an autoregressive temporal autocorrelation structure of order 1, i.e., $AR(1)$, at two spatial resolutions. The model equations below for 3a and 3b hold for all $k$ and $t$. Model 3a included an autocorrelated normally distributed error term $\eta_t$, with $\eta = (\eta_1, . . . , \eta_T )$ and a global temporal autocorrelation parameter, $\rho$, and variance $\sigma^2_\eta$, where $\rho$ is given a uniform prior distribution between 0 and 1, (again, non-negative) and the remaining model parameters are given the same prior distributions as before. 
\begin{subequations}
\begin{flalign}
Y_{kt} | \mu_{kt} &\sim \text{Pois}(\mu_{kt}) \\
\ln(\mu_{kt}) &= \sum_{i=0}^{p} x_{kti} \beta_i + O_{kt} + \Phi_{k} + \eta_t  \nonumber\\
&\Phi|\Sigma \sim \text{MVN}(0,\Sigma)\medspace\nonumber\\
& \qquad \Sigma = \sigma_\Phi^2 (D - \lambda W)^{-1} \nonumber\\
& \eta_t|\rho,\eta_{t-1}, \sigma_\eta^2 \sim \text{N}(\rho\eta_{t-1}, \sigma_\eta^2) \quad \text{for all } t=2,3,...,T \nonumber \\
& \qquad \beta_i \sim N(0,100) \nonumber\\
& \qquad \lambda, \rho \sim \text{U}(0, 1)\nonumber\\
& \qquad  \sigma_\Phi^2, \sigma_\eta^2 \sim \text{inverse-Gamma}(1,1)  \nonumber 
\end{flalign}
In contrast, Model 3b stipulated tributary-specific temporal autocorrelation for $g=1,2,3$: 
\begin{flalign}
Y_{ktg} | \mu_{ktg} &\sim \text{Pois}(\mu_{ktg})  \\
\ln(\mu_{ktg}) &= \sum_{i=0}^{p} x_{kti} \beta_i + O_{kt} + \Phi_{k} + \eta_{gt}  \nonumber\\
&\Phi|\Sigma \sim \text{MVN}(0,\Sigma)\medspace\nonumber\\
& \qquad \Sigma = \sigma_\Phi^2 (D - \lambda W)^{-1} \nonumber\\
& \eta_{gt}|\rho_g,\eta_{g,t-1},\sigma_\eta^{2} \sim \text{N}(\rho_g\eta_{g,t-1}, \sigma_\eta^2) \quad \text{for all } t=2,3,...,T \nonumber \\
& \qquad \text{logit}(\rho_{g}) = \text{logit}(P) + r_g \nonumber\\
& \qquad r_3 = - r_1 - r_2 \nonumber\\
& \qquad \qquad\beta_i \sim N(0,100) \nonumber\\
& \qquad \qquad \lambda, P \sim \text{U}(0, 1)\nonumber\\
& \qquad \qquad  \sigma_\Phi^2, \sigma_\eta^2 \sim \text{inverse-Gamma}(1,1)  \nonumber \\
& \qquad \qquad  r_1, r_2  \sim \text{N}(0,0.25) \nonumber
\end{flalign}
\end{subequations}

Here, $\eta_{gt}$ is the normally distributed AR(1) error term for year $t$ and tributary $g$, with a local temporal autocorrelation parameter $\rho_g$ and global variance $\sigma^2_\eta$. the complete set is denoted by $\eta=(\eta_1,\eta_2,\eta_3$), where $\eta_g=(\eta_{g1},\eta_{g2},...,\eta_{gT})$.  Here, each $\rho_g$ on the logit scale is modeled as the logit of a global temporal autocorrelation parameter $P$ plus a tributary-specific offset $r_g$, subject to the sum-to-zero constraint $\sum_{i=1}^{3} r_{g} = 0$. Two of the $r_g$s are given normal priors of $N(0,0.25)$ which reflect a compromise between the lack of pre-existing knowledge about these parameters and a desire to constrain the distributions from unrealistic extremes \citep[][and Fig. \ref{fig:r}]{gelman2013bayesian}. The inverse-logit transformation, $\text{logit}^{-1}(u)=\frac{e^{u}}{1+e^{u}}$ for any real number $u$ (here, $u=\text{logit}(\rho_g)$), constrains $\rho_g$ between 0 and 1. Similarly, $P$ is given a uniform prior between 0 and 1. The remaining model parameters were given the same prior distributions as before.
\subsection{Model 4}\label{meth:model 4}
For the final model, we considered a non-separable spatiotemporal random effect. The spatiotemporal structure includes a multivariate first-order autoregressive process with a first-order spatial CAR structure. The data level and linear predictor of the resulting hierarchical model are:


\begin{align}
Y_{kt}|\mu_{kt} &\sim \text{Pois}(\mu_{kt}) &\textrm{and}& &\ln(\mu_{kt}) = \sum_{i=0}^{p} x_{kti} \beta_i + O_{kt} + \Phi_{kt}.\nonumber
\end{align}
Here, the $\Phi_{kt}$ term is the random effect associated with section $k$ in year $t$, with the complete set denoted by $\Phi = (\Phi_1, . . . , \Phi_T )$, where each $\Phi_t = (\Phi_{1t}, . . . , \Phi_{Kt})$ is the $t$th map of spatial random effects.

\begin{align}
&\Phi_1|\Sigma \sim \text{MVN}(0,\Sigma)\medspace\medspace \\
& \qquad \Sigma = \sigma_\Phi^2 (D - \lambda W)^{-1} \nonumber\\
&\Phi_t|\rho,\Phi_{t-1},\Sigma \sim \text{MVN}(\rho\Phi_{t-1},\Sigma),\medspace \text{when} \medspace t > 1 \nonumber \\
& \qquad \Sigma = \sigma_\Phi^2 (D - \lambda W)^{-1} \nonumber\\
& \qquad \qquad \beta_i \sim N(0,100) \nonumber \\
& \qquad \qquad \sigma_\Phi^2 \sim \text{inverse-Gamma(1,1)}  \nonumber\\
& \qquad \qquad \lambda, \rho \sim \text{U}(0, 1). \nonumber
\end{align}

The spatiotemporal autocorrelation structure is stipulated by replacing $\eta_t$ in Model 3a with the entire $\Phi_t$ map, an approach employed by \citet{rushworth2014spatio}, and represents the spatiotemporal pattern in the mean response with a single set of spatially and temporally autocorrelated random effects. The $\Phi_t$ map follows a multivariate autoregressive process of order one. Thus, in year $t = 1$, the $\Phi_1$ map assumes a strictly CAR structure. However, when $t > 1$, temporal autocorrelation is induced by explicitly allowing $\Phi_t$ to have conditional mean equal to $\rho\Phi_{t-1}$.

The regression coefficients, autocorrelation parameters, and variance parameters were given the same prior distributions as before.

\section{Model Implementation and Validation} \label{sec:validate}
For each model, Bayesian inference required numerical approximation of the joint posterior distribution of all model parameters including the vectors of random effects $\theta, \Phi$, and $\eta$. To this end, we implemented the above models using the Stan programming language for Bayesian inference to generate Markov chain Monte Carlo (MCMC) samples from the posterior \citep{gelman2015stan}. For each model, we ran four parallel Markov chains, each with 15,000 iterations for the warm-up/adaptive phase (and subsequently discarded as burn-in), and another 15,000 iterations as posterior samples (i.e., 60,000 draws in total for posterior inference). Convergence of the chains was determined both by visual inspection of trace plots (e.g., Fig. \ref{fig:Traceplot}) and through inspection of the split $\hat{R}$ statistic. All sampled parameters had an $\hat{R}$ value less than 1.01, indicating chain convergence \citep{gelman2015stan}. Covariates and interactions whose regression coefficients had credible intervals (CIs) that excluded 0 at a credible level of 80\% (i.e., reasonably high for hierarchical Bayesian inference) were considered scientifically relevant to juvenile blue crab abundance. All CIs referenced here are highest posterior density intervals (HPDIs) \citep{mcelreath2018statistical}.

Model validation and relative predictive performance were assessed using out-of-sample cross validation (CV), whereby a subset of the full data was used to train models, and the trained models were then used to predict the withheld data. Given the spatiotemporal dependence structures within our model, common CV procedures, such as leave-one-out (LOO), are difficult to interpret if the goal is to assess predictive performance, because withheld observations depend on other observations from different time periods and different spatial units in addition to the dependence on the model parameters \citep{burkner2021efficient}. For example, withholding random observations in time-series models will still allow information from the future to influence predictions of the past. Instead, we employed the leave-future-out (LFO) CV approach to evaluate predictive performance through withholding future samples \citep{burkner2020approximate}. Thus, prior to CV analysis, the data from the final year of the study, 2017, were excluded from the models. Then, the above Models 1--4 were fitted to the reduced dataset for both model inference (whose results appear under Section \ref{sec:results}) as well as CV. For CV, 80\% Bayesian prediction intervals were computed from the posterior predictive distributions of the excluded values, as a forecasting exercise. The final step of CV analysis was to compare the excluded blue crab count values to the forecast prediction intervals. Note that due to missing data in one of the sections in 2017 (see Section \ref{sec:datasummary}), CV was only possible for $n=36$ sections.

\section{Results}\label{sec:results}
\subsection{Data Summary}\label{sec:datasummary}
In total, 75,103 juvenile blue crabs between 20-40 mm CW were captured between 1996 and 2017 from April to December in the York, James, and Rappahannock Rivers. The highest abundances of juvenile blue crabs occurred in upriver locations of each tributary (Fig. \ref{fig:importance2}a, \ref{fig:importance3}a). Relative seagrass area was highest in the York River and lowest in the James River (Table \ref{tab:Summary}). Relative marsh area and turbidity were highest in the York River and lowest in the Rappahannock River (Table \ref{tab:Summary}). Within each tributary, turbidity generally increased with distance upriver.

\begin{figure}[htbp]
\center{\includegraphics[width=\textwidth]{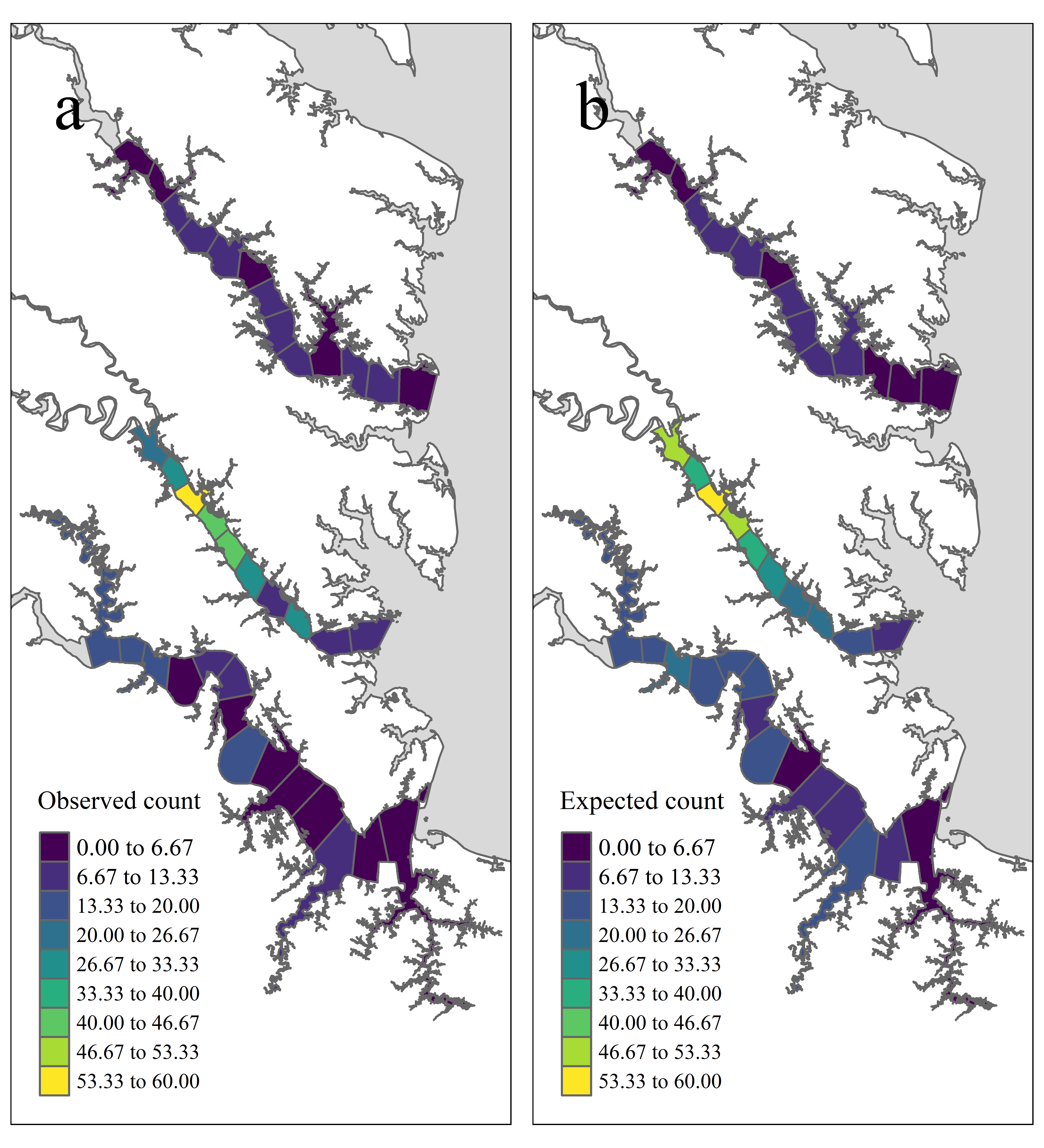}}
\caption{Temporally aggregated observed and expected juvenile blue crab abundance in each tributary section based on inter-annual grand means of model quantities from years 2009-2017 and management after 2008 (see Section \ref{sec:prioritized} for definitions). Panel (\textbf{a}) shows the mean observed juvenile blue crab abundance ($\overline{y}_{k+}$), while Panel (\textbf{b}) shows the pseudo-posterior median of the expected abundance on the count scale ($\overline{\mu}_{k+}$).}
\label{fig:importance2}
\end{figure}

\begin{figure}[htbp]
\center{\includegraphics[width=\textwidth]{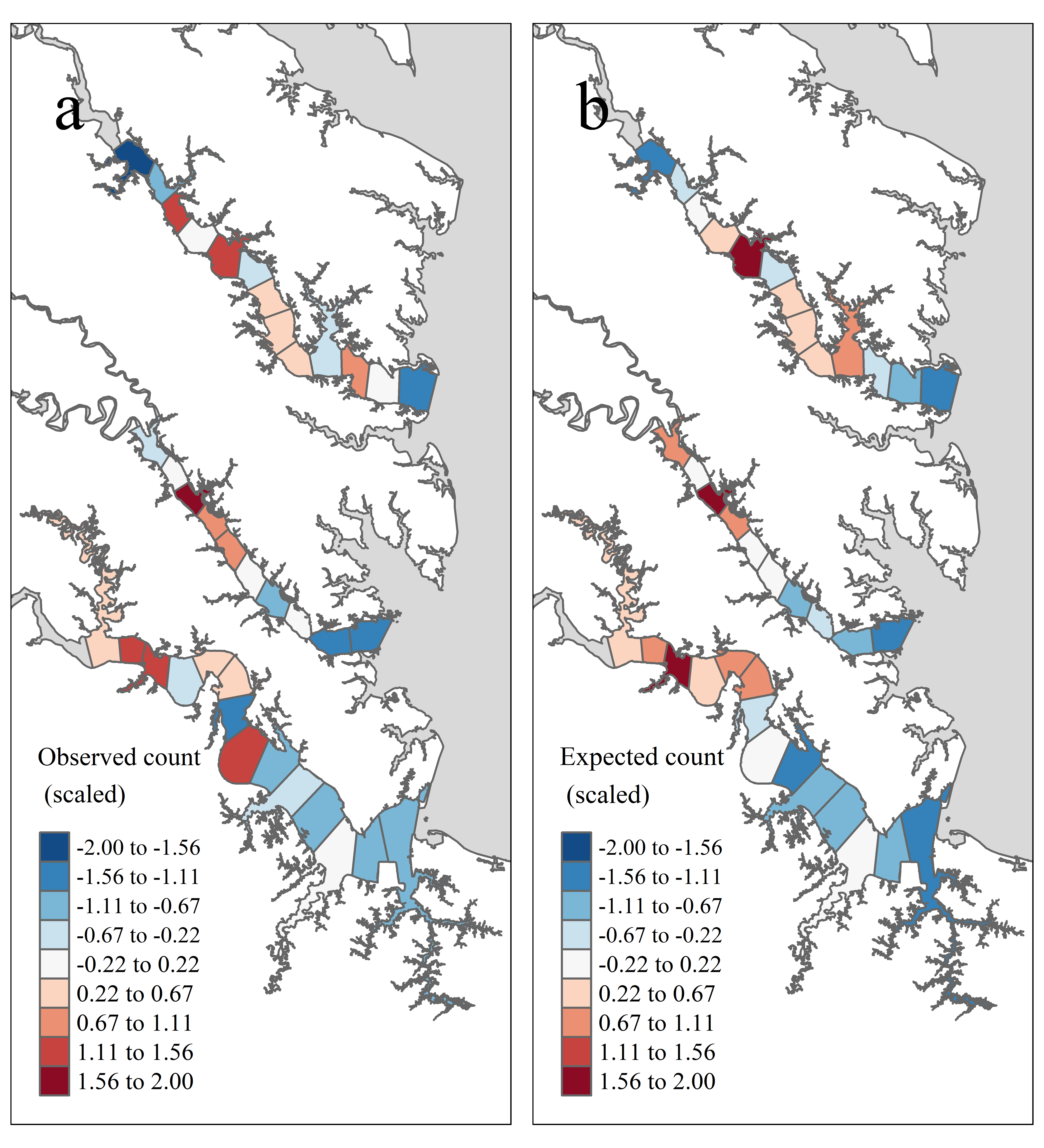}}
\caption{Temporally aggregated observed and expected juvenile blue crab abundance in each tributary section based on inter-annual grand means of model quantities from years 2009-2017 and management after 2008 (see Section \ref{sec:prioritized} for definitions), standardized within tributary. Panel (\textbf{a}) shows the tributary-specific standardized values of $\overline{y}_{k+}$ (mean observed juvenile blue crab abundance), while Panel (\textbf{b}) shows the tributary-specific standardized values of $\overline{\mu}_{k+}$ (pseudo-posterior median of the expected abundance on the count scale).}
\label{fig:importance3}
\end{figure}

\begin{table}[ht]
\caption{Mean (minimum--maximum) section-year values for Secchi disk depth, salinity, relative seagrass area (RSA), relative marsh area (RMA), and predator abundance for each tributary. Values were calculated from data collected over a 22-year period (1996--2017). }
\begin{center}
\begin{tabular}{lccccc}
  \hline
 Tributary & Secchi & Salinity & RSA & RMA & Predator abundance \\ 
  \hline
 James & 0.77 & 11.13 & 0 & 0.15 & 287.52 \\ 
 & (0.29--1.59) & (0.5--22.12) & (0--0.08) & (0.02--0.36) & (0--2838) \\ 
 Rappahannock & 1.1 & 12.74 & 0.01 & 0.08 & 147.32 \\ 
 & (0.26--2.34) & (2.73--19.39) & (0--0.09) & (0.01--0.38) & (0--2154) \\ 
 York & 0.78 & 15.85 & 0.02 & 0.23 & 383.62 \\ 
 & (0.38--1.57) & (6.81--22.24) & (0--0.17) & (0.01--0.48) & (2--2381) \\ 
   \hline
\end{tabular}
\end{center}
\label{tab:Summary}
\end{table}

\subsection{Model Selection and Validation}
Cross validation indicated that the non-separable spatiotemporal model, \nameref{meth:model 4}, best described patterns in juvenile blue crab abundance. The 80\% posterior prediction intervals from \nameref{meth:model 4} contained 81.1\% of withheld 2017 data, while \nameref{meth:model 1} (random effect only), \nameref{meth:model 2} (spatial-only CAR model) , \nameref{meth:model 3}a (spatial CAR model with separable, global AR(1) term), and \nameref{meth:model 3}b (spatial CAR model with separable, tributary-specific AR(1) term) captured 21.6, 18.9, 43.2, and 70.3\% of withheld data, respectively (Fig. \ref{fig:CV}). A full description of model validation and predictive performance is provided in Appendix \ref{appendix:validation}. Hereafter, inferences are made from \nameref{meth:model 4} only.

\begin{figure}[htbp]
\center{\includegraphics[width=\textwidth]{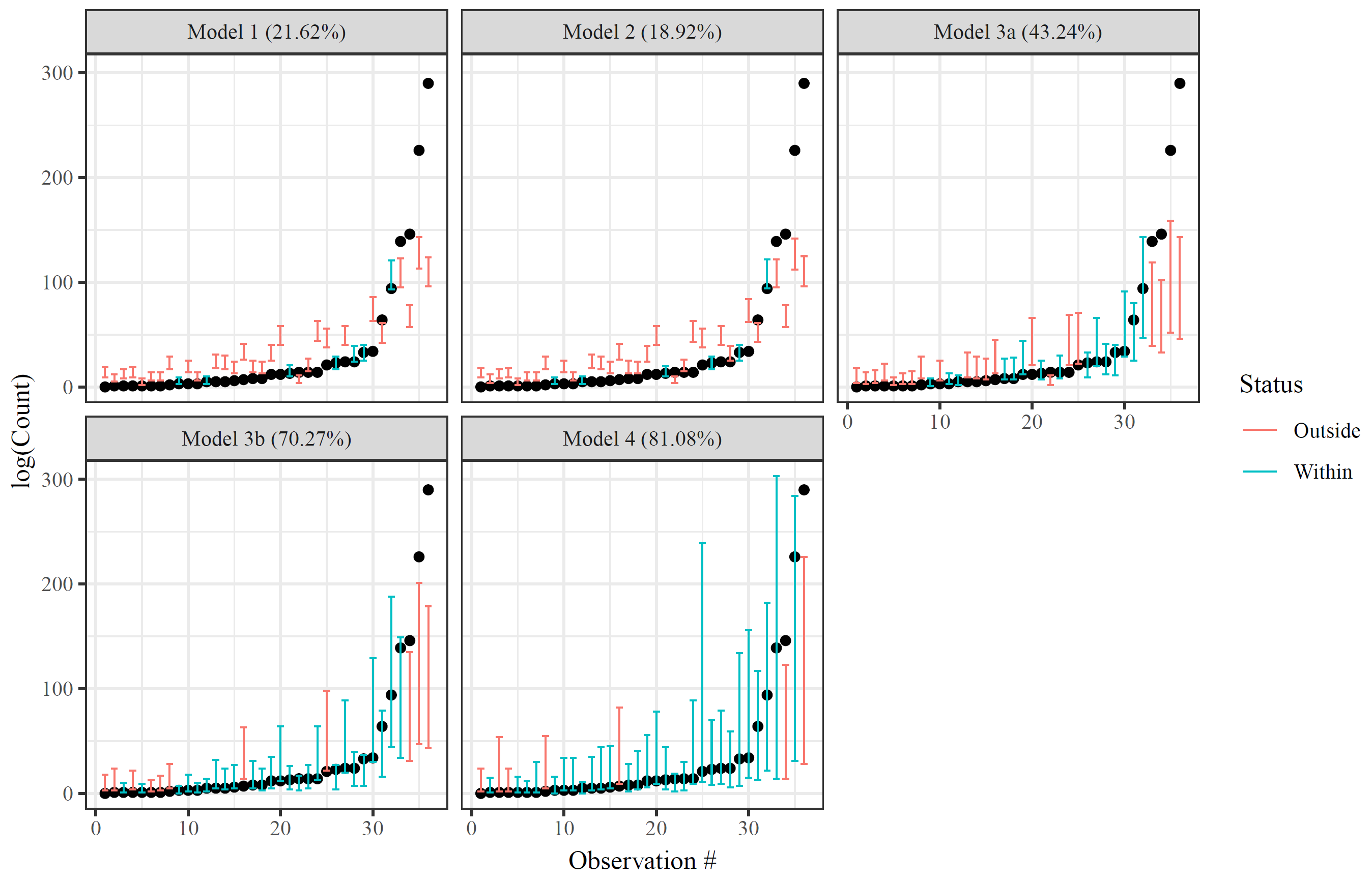}}
\caption{Leave-future-out cross validation results for Models 1--4. Bars denote nominal 80\% Bayesian prediction intervals derived from posterior predictive distributions, while dots depict observed crab counts for 2017. Red bars indicate an observed value is outside the prediction interval, while blue bars indicate an observed value is within the  prediction interval. Actual coverage percentages of prediction intervals (= \% of blue) appear in the panel headings.}
\label{fig:CV}
\end{figure}

\subsection{Implications of Prioritized Areas for Conservation}\label{sec:prioritized}
Using posterior distributions derived from \nameref{meth:model 4}, we aggregated over time and made spatial-only predictions of juvenile blue crab abundance to identify areas of high abundance. For continuous predictors, data were aggregated for each section over 2009--2017 to obtain inter-annual grand means, i.e., $\overline{x}_{k+ i} = \sum_{t=1}^{T} x_{kti}/T$ for continuous predictor variable $x_{kti}$, where $T=9$ for all sections except $T=8$ for the section with no trawl data in 2017. The same aggregation was applied respectively to the log-transformed tow distance offset term $O_{kt}$ and each $d$th posterior draw for the spatiotemporal random-effect term $\Phi^{(d)}_{kt}$ to define $\overline{O}_{k+}$ and $\overline{\Phi}^{(d)}_{k+}$. Thus, for abundance, $\overline{\mu}^{(d)}_{k+}$ denotes a temporally aggregated posterior draw of the expected abundance from replacing $\mu_{kt}$ of \nameref{meth:model 4} with $\overline{\mu}^{(d)}_{k+}$ that was computed using $\overline{x}_{k+ i}, \overline{O}_{k+}$, and $\overline{\Phi}^{(d)}_{k+}$; the set $\{\overline{\mu}^{(1)}_{k+},\overline{\mu}^{(2)}_{k+},...\overline{\mu}^{(60000)}_{k+}\}$ for each $k$ forms a \textit{pseudo-posterior distribution} of the temporally aggregated expected abundance $\overline{\mu}_{k+}$ for spatial section $k$. (A true posterior distribution would require fitting a spatial-only version of \nameref{meth:model 4} that directly models temporally averaged counts $\overline{y}_{k+}$.) We limited spatial-only comparisons to 2009--2017 due to the change in management following 2008, which was a  categorical predictor and could not be reasonably averaged over time. In addition to inspecting the $n=37$ values of pseudo-posterior median for $\overline{\mu}_{k+}$ ($k=1,...,n$), we computed tributary-specific standardized values (centered and scaled to have unit variance) of the pseudo-posterior medians of $\overline{\mu}_{k+}$ to determine regions of locally high and low abundance relative to each tributary. Resulting sections with predictions corresponding to higher average relative juvenile blue crab abundance were interpreted as more productive within tributary, whereas sections with predictions corresponding to lower average relative juvenile blue crab abundance were interpreted as less productive within tributary.

According to pseudo-posterior medians of $\overline{\mu}_{k+}$, upriver sections of tributaries consistently harbored highest crab abundances (Figs. \ref{fig:importance2}b and \ref{fig:importance3}b). In particular, upriver sections in the York River were very high, with a pseudo-posterior median of 30--60 crabs per 1000 m towed. Upriver sections in the James River had a pseudo-posterior median of 13--26 crabs per 1000 m towed, whereas those in the Rappahannock River were much lower at 0--13 crabs per 1000 m towed. Pseudo-posterior medians for $\overline{\mu}_{k+}$ were generally consistent with observed juvenile blue crab abundances $\overline{y}_{k+}$ in each section from 2009--2017 (Figs. \ref{fig:importance2} and \ref{fig:importance3}).

\subsection{Drivers of Juvenile Blue Crab Abundance}\label{sec:drivers}
Based on posterior distributions of Model 4 parameters, tributary, turbidity, relative marsh area, and relative marsh area $\times$ turbidity were relevant drivers of juvenile blue crab abundance. All tributaries differed in average juvenile blue crab abundances, with posterior probabilities $P(\beta_{\text{York}} > \beta_{\text{James}}| \text{data}) > 0.99$, $P(\beta_{\text{York}} > \beta_{\text{Rappahannock}}| \text{data}) > 0.99$, and $P(\beta_{\text{James}} > \beta_{\text{Rappahannock}}| \text{data}) = 0.99$. Turbidity, marsh, and their interaction positively influenced juvenile blue crab abundance --- posterior medians (and 80\% CIs) were: $\beta_{\text{Turbidity}}$ = 0.48 (0.20--0.77), $\beta_{\text{Marsh}}$ = 2.55 (1.07--4.01), and $\beta_{\text{Marsh x Turbidity}}$ = 3.42 (1.34--5.49). Regression coefficients $\beta_{\text{Seagrass}}$, $\beta_{\text{Predator}}$, and $\beta_{\text{Management}}$ had respective 80\% CIs that included 0. Supporting posterior summaries and graphics are in Table \ref{table:posterior} and Fig. \ref{fig:Posterior1}.

\begin{table}[ht]
\caption{Posterior summary statistics (median and 80\% CI's) of regression coefficients $\beta$ as well as autocorrelation parameters $\lambda$ (spatial) and $\rho$ (temporal) from Model 4. For regression coefficients, the symbol ``*'' indicates the 80\% CI does not contain 0}
\begin{center}
\begin{tabular}{lccc}
  \hline
 Parameter &  10\% & 50\% & 90\%  \\ 
  \hline
  $\beta_0$*  & -4.66 & -4.33 & -3.99 \\ 
  $\beta_{\text{Turbidity}}$*  & 0.20 & 0.48 & 0.77 \\ 
  $\beta_{\text{Seagrass}}$ & -4.44 & -1.24 & 1.88 \\ 
  $\beta_{\text{Marsh}}$ * & 1.07 & 2.55 & 4.01 \\
  $\beta_{\text{Marsh} \times \text{Turbidity}}$* & 1.34 & 3.42 & 5.49 \\ 
  $\beta_{\text{Predator}}$ & -0.00 & 0.04 & 0.09 \\ 
  $\beta_{\text{Management}}$ & -0.08 & 0.09 & 0.27 \\ 
  $\beta_{\text{Rappahannock}}$* & -0.56 & -0.34 & -0.13 \\ 
  $\beta_{\text{York}}$* & 0.53 & 0.74 & 0.97 \\ 
  $\lambda$ & 0.48 & 0.55 & 0.61 \\ 
  $\rho$ & 0.05 & 0.11 & 0.16 \\
   \hline
\end{tabular}
\end{center}
\label{table:posterior}
\end{table}

Conditional effects plots (Figs. \ref{fig:interaction1} and \ref{fig:interaction2}) were used to visualize the relationship between juvenile blue crab abundance and predictors relative marsh area and turbidity.  For conditional effects plots, we considered the function $\mu_{\text{cond}}(x_{\text{Turbidity}},x_{\text{Marsh}})=x_{\text{Turbidity}}\beta_{\text{Turbidity}} +x_{\text{Marsh}}\beta_{\text{Marsh}} + x_{\text{Marsh}} x_{\text{Turbidity}}\beta_{\text{Marsh}\times\text{Turbidity}}+\beta_{\text{Management}}+\ln(1000)$, which re-expresses the juvenile blue crab expected abundance $\mu_{kt}$ as a function of $x_{\text{Turbidity}}$ and $x_{\text{Marsh}}$ as the only varying predictors, while all other continuous predictor variables were held at 0 and the tow offset term was held at 1000 m, while categorical variables were held at the James River (tributary) and post 2009 period. The relationship between each varying predictor and $\mu_{\text{cond}}$ was plotted (along with credible bands) with the other varying predictor held at fixed percentiles (1, 20, 40, 60, 80, and 99\%) to visualize interaction effects. Relative marsh area influenced juvenile blue crab abundance negatively at low turbidities (i.e., $\le -$0.81 = median) and positively at high turbidities (i.e., $\ge -$0.81 = median) (Fig. \ref{fig:interaction1}). In contrast, turbidity influenced crab abundance positively at both low and high relative marsh area values, with the strength of the relationship between turbidity and abundance growing progressively stronger at high relative marsh area values (Fig. \ref{fig:interaction2}).  

\begin{figure}[htbp]
\center{\includegraphics[width=\textwidth]{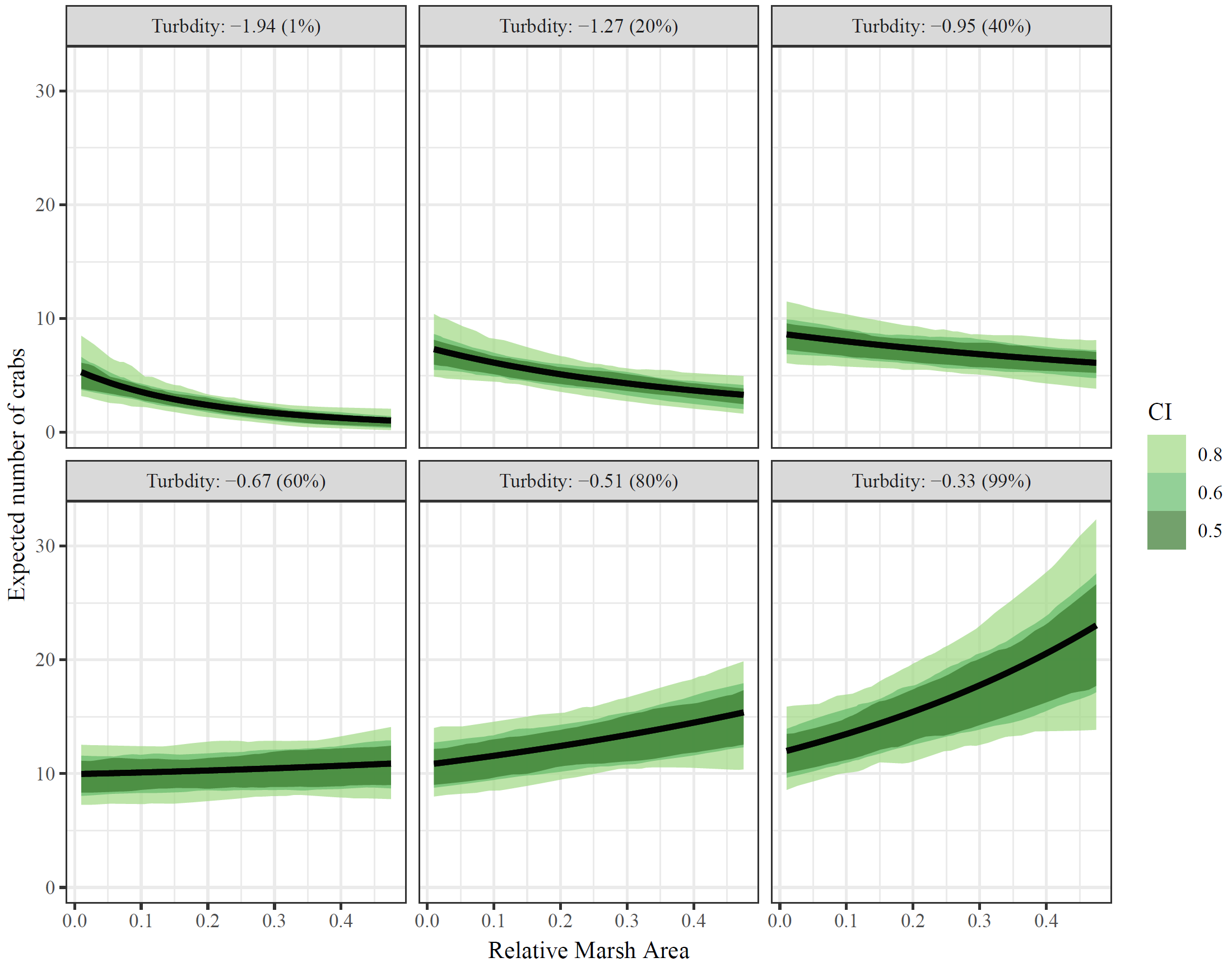}}
\caption{Conditional effects plots depicting relationship between juvenile blue crab abundance per 1000 m towed ($\mu_{\text{cond}}$) vs relative marsh area (RMA) at turbidity values corresponding to 1, 20, 40, 60, 80, and 99\% percentiles to visualize interaction effects between relative marsh area and turbidity on crab abundance. All other continuous variables were held at 0 and categorical variables at the James River (tributary) and post 2008 (management). Colored bands indicate credible bands of $\mu_{\text{cond}}$. (See Section \ref{sec:drivers} for definitions.)} 
\label{fig:interaction1}
\end{figure}

\begin{figure}[htbp]
\center{\includegraphics[width=\textwidth]{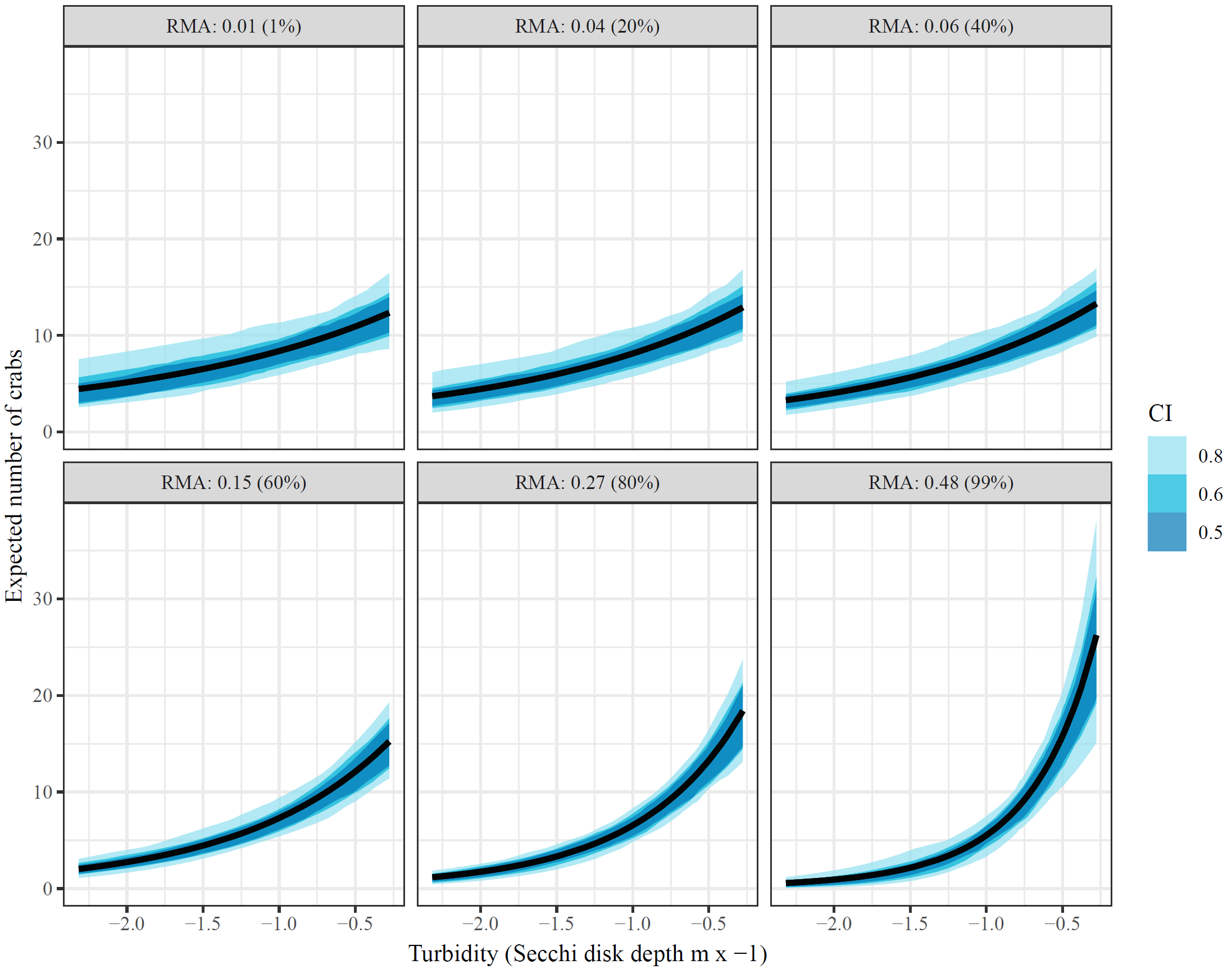}}
\caption{Conditional effects plots depicting relationship between juvenile blue crab abundance per 1000 m towed ($\mu_{\text{cond}}$) vs turbidity at relative marsh area (RMA) values corresponding to 1, 20, 40, 60, 80, and 99\% percentiles to visualize interaction effects between relative marsh area and turbidity on crab abundance. All other continuous variables were held at 0 and categorical variables at the James River (tributary) and post 2008 (management). Colored bands indicate credible bands of $\mu_{\text{cond}}$. (See Section \ref{sec:drivers} for definitions.)}
\label{fig:interaction2}
\end{figure}

\subsection{Spatiotemporal Dependence}
The posterior distribution of $\lambda$ indicated that substantial spatial dependence existed within the data (Fig. \ref{fig:Posterior2}). Posterior distributions of $\lambda$ and $\rho$ yielded medians (80\% CIs) of 0.61 (0.55--0.68) and 0.14 (0.08--0.19), respectively. Although the magnitude of $\rho$ was small, this non-separable spatiotemporal model, when compared to simpler models, gave leave-future-out 80\% Bayesian prediction intervals that had the highest coverage of the withheld 2017 data, and the coverage was close to its nominal 80\% (Fig. \ref{fig:CV}). Moreover, among the competing models the spatiotemporal structure was strong enough that posteriors of the fixed‐effect coefficients changed markedly when spatially and spatiotemporally structured random effects were included (Fig. \ref{fig:Posterior3}).

\section{Discussion}
Abundance of juvenile blue crabs varied spatially both within and among the three tributaries, James, York and Rappahannock Rivers. Within all tributaries, abundance of juvenile blue crabs consistently peaked in upriver sections. Given the limited mobility of juvenile blue crabs $<$60 mm CW, we interpret high 20--40 mm CW abundance in upriver areas as reflective of highly productive nursery habitats, as previously hypothesized for the York River \citep{lipcius2005density,seitz2005food}. Moreover, juvenile blue crab abundance was associated with specific environmental characteristics, especially with high turbidity and extensive marsh area near the turbidity maximum of each tributary. These findings offer an initial quantification of multiple environmental components of highly productive nursery locations within the seascape paradigm for juvenile blue crabs in lower Chesapeake Bay. 

\subsection{Environmental Determinants of Juvenile Blue Crab Abundance}
Availability of marsh habitat and high turbidity were the most important predictors of juvenile blue crab abundance, which was strongly and positively related to turbidity, and increased with the availability of salt marsh habitat relative to geographic area. However, the substantial interaction between marsh habitat and turbidity required that inferences on the relationship between marsh habitat or turbidity and juvenile blue crab abundance be made within the context of the other factor.

In areas characterized by low turbidity (i.e., mean Secchi depth $>$1 m), the effect of marsh habitat ranged from negligible to negative. Conversely, in locations of high turbidity (i.e., mean Secchi depth $<$1 m), juvenile blue crab abundance was positively associated with availability of marsh habitat. About half of the section-years considered in our study were characterized by high turbidity where marsh availability was positively related to crab abundance. Turbidity and crab abundance were always related positively, and this relationship grew stronger (steeper slope) as marsh area increased in a river section. 

While relative area of adjacent marsh habitat was positively related to juvenile blue crab abundance, other potential nursery habitats were weakly associated with crab abundance. Specifically, relative seagrass area was not associated with juvenile crab abundance. This was particularly surprising for seagrass, which has long been considered the preferred nursery habitat for small juvenile blue crabs \citep{orth1987utilization, lipcius2007post}. We propose that the lack of association between juvenile blue crab abundance and these habitat types reflects differences between nursery habitat contributions per unit area \citep[proposed by][]{beck2001identification} versus effective nursery habitat and total contribution to the adult segment of the population \citep[proposed by][]{dahlgren2006marine}. At the tributary spatial scale of our study, the areal extent of marsh habitat relative to the area of river sections was much greater than that of seagrass meadows, particularly in the York and James Rivers. Moreover, section-years harboring seagrass meadows (such as in downriver York and midriver-downriver Rappahannock sections) were not associated with high juvenile blue crab abundance. Consequently, at the tributary scale, the potentially high production of juvenile blue crabs per unit area expected in river sections with seagrass meadows was likely obfuscated by the broad areal extent of marsh habitat in other sections. Furthermore, in downriver sections where seagrass was present, a substantial fraction of juveniles 20--40 mm CW may have remained in seagrass where they were not susceptible to capture by the trawl \citep{orth1987utilization, lipcius2005density, seitz2005food, ralph2013broad}.

Predator abundance was not related to juvenile blue crab abundance. The apparent lack of an effect of predator abundance on juvenile blue crab abundance may reflect high refuge capacity of crab nurseries or increased availability of alternative prey in locations harboring high blue crab abundance \citep{lipcius2005density}. Moreover, finfish predators are highly mobile and not likely to remain in a specific section. Regardless, our findings suggest that abundances of juvenile blue crabs at the regional scale are largely driven by bottom-up controls rather than top-down controls, which is consistent with studies of blue crab abundance in highly turbid, upriver localities harboring expansive marsh habitat \citep{seitz2003potential, posey2005importance, seitz2005food}. 

Finally, juvenile blue crab abundance differed substantially among the three tributaries. These spatial patterns in abundance likely reflected tributary-specific characteristics that we did not consider in our models (e.g., differences in flow, bathymetry, total area, geographic position relative to the mouth of the Chesapeake Bay, or land-use patterns). Ultimately, spatial variation in juvenile blue crab abundance among tributaries indicates that tributaries in the Chesapeake Bay are not equal as nursery areas for the blue crab population, and that further studies should quantify tributary-specific production to the population. 

\subsection{Effect of Management}
Changes in management of the blue crab population in the Chesapeake Bay after 2009 were positively associated with juvenile blue crab abundance, but not strongly, in contrast to the findings of other studies \citep{MDNR2019, Dredge2020}. One explanation for this result is the potential effect of cannibalism by larger juveniles and adults on small juveniles, which might negate positive effects of increased recruitment from a larger spawning stock \citep{lipcius2002concurrent}. More likely, the effect of management may depend upon specific habitats, especially those where juveniles are abundant, such as in habitats with expansive marshes. Other sections where juveniles are not as abundant, such as unvegetated habitat, may not be able to support higher levels of recruitment, which would confound singular interpretation of management. Targeted analyses of the effects of management in specific habitats are ongoing to resolve this issue.

\subsection{Prioritized Areas for Conservation}
An objective of this study was to assist management to prioritize and direct restoration and conservation efforts of the blue crab within Chesapeake Bay as well as other blue crab stocks along its geographic range. Although previous focus of blue crab nursery studies was on seagrass meadows \citep{Ralph2014}, salt marshes and certain unstructured, high turbidity habitats appear more valuable at the tributary and regional scales due to their extensive areal cover. Our best fitting model indicates that expansive salt marshes in highly turbid upriver locations are highly productive nurseries for this ecologically and economically exploited species. As a result, a major recommendation of this paper is the inclusion of these habitats in future conservation targets.

\subsection{Relevance}
The EFH provisions of the Magnuson-Stevens Act directs fishery management councils to utilize the best available science to describe and identify EFH for federally managed species and protect them to the extent practicable \citep{MSA2007}. The highest level of EFH information is level 4: production rates by habitat type; yet level 4 EFH information is largely unavailable for most commercially harvested species, particularly at spatial and temporal scales needed for effective fisheries management. This lack of level 4 EFH information is currently limiting the inclusion of habitat effects in stock assessments and in ecosystem-based fisheries management plans \citep{gruss2017recommendations}. Furthermore, area-based estimates of nursery habitat value may inform decision-making related to protected area management and habitat restoration, by allowing the per unit area contribution of protected or restored habitat to be quantified \citep{zu2021estimating}.

Understanding the relative contribution of both structured habitats and other environmental factors on the productivity of a given area is important, as many conditions resulting in such productivity are diminishing. Some structured nursery habitats are declining, especially \textit{Z. marina} eelgrass beds due to direct and indirect anthropogenic influences such as land-use change and long-term warming of Chesapeake Bay \citep{orth2010long,moore2014impacts, patrick2018land}. Similarly, salt marshes have been reduced by coastal development and shoreline hardening \citep{silliman2009human}. 

Scientists and managers have generally assumed that when structured habitats are degraded, the services they provide such as nursery habitat for valuable marine species are lost \citep{peterson2003conceptual}. Therefore, state and federal agencies have long invested in coastal habitat conservation and restoration to recover lost production. However, these investments have often preceded the availability of, and thus would be enhanced by the development of, rigorous analytical tools capable of quantifying the ecosystem services expected from conservation actions and habitat restoration efforts. While standalone small-scale studies have been and remain important tools to initially assess nursery value of structured habitats and other environmental factors, targeted comprehensive applications of survey data collected over broad spatial and temporal scales are a vital complement to generalize inference of nursery function, highlight highly productive regions, and inform regional management strategies.

\begin{appendix}
\section{More on predictor variables}
\label{appendix:parameters}
Unstructured habitat constitutes the majority of available shallow habitat in Chesapeake Bay, but varies considerably in food availability and predation refuge \cite{lipcius2005density}. Evidence suggested unstructured mud may serve as an alternative nursery for juveniles where structurally complex habitat is unavailable due to relatively abundant alternative prey and potential for juveniles to bury deep in the soft substrate \cite{mense1989distribution,rakocinski2003soft}. Thus, in the earliest exploratory models, we had included mean percent mud composition of substrates in each section-year as a continuous covariate, in addition to those presented in Table \ref{table:description}. However, 80\% credible intervals for the corresponding regression coefficient of this variable consistently included 0, and inclusion of the variable did not otherwise change inference results of the initial models. In contrast, the other variables, such as seagrass, management status, and predator abundance, were always kept in our models regardless of their statistical importance in explaining juvenile blue crab abundance, due to their implications on large-scale blue crab population management. Because percent mud composition did not carry the same implications with respect to management, it was excluded as a variable of interest in all models presented in this article.

\section{Defining areal units}
\label{appendix:areal units}
Note that despite the arbitrary nature of areal unit definitions in practice, the one we employed in our work here did not meaningfully influence our results, or bias, our inference. In fact, initially, we explored numerous areal unit configurations when aggregating spatially random trawls. Alternative configurations included dividing each tributary into i) ten sections whose lengths were tributary dependent, ii) sections based on morphologically meaningful characteristics (e.g., branching structures and choke points), and iii) sections $\sim$2km in length along the tributary axis. In all cases, parameter estimates from models were practically identical. The final areal unit configuration was chosen based on the high number of areal units per year produced, and only a single section-year had 0 trawl tows.

\section{Model validation and predictive performance}
\label{appendix:validation}
Cross validation (CV) is a robust, generic method to adjudicate between competing statistical models. Unlike information theoretic criteria (e.g., AIC, BIC, DIC), cross validation assesses predictive performance directly by separating the data in a part that is used for fitting (i.e., training set) and another used to assess predictive adequacy (i.e., test set). Cross validation preference goes to the model that best predicts the out-of-sample test set withheld. 

Cross validation is helpful in determining relative model  generalizability. In a Bayesian CV framework, prediction intervals are computed using the posterior predictive distributions of the excluded values in the test set based on posterior distributions of model parameters to simulate the training set. Generalizability is determined based on the observed coverage, i.e., the proportion of excluded values which are successfully captured by their respective prediction intervals.

We used 80\% prediction intervals to infer model performance of our suite of candidate models. Models yielding an observed coverage differing greatly from the nominal Bayesian predictive credible level of 80\% may indicate underfitting/overfitting. Cross validation results from Models 1, 2, 3a, and 3b, all indicated underfitting (being less complex than Model 4) and poor predictive performance. In contrast, posterior prediction intervals of Model 4 contained 81\% of excluded data ($n=36$), indicating overall superior predictive performance relative to all other candidate models. Hence, we selected Model 4 as the model which best represents our observed data as well as the most generalizable model.

In contrast with simpler models, Model 4 is characterized by greater uncertainty in posterior distributions of predictor coefficients as well as posterior predictive credible intervals used in cross validation (Figs. \ref{fig:CV} and \ref{fig:Posterior3}). This is a frequent characteristic of models with increasing complexity. Complex models (with a larger number of unknown model parameters) lead to more uncertainty in the inference, whereas simpler models which are inadequate in capturing latent dependence processes would give incorrect inference, irrespective of the amount of uncertainty. 
 \newpage
 
 \SupplementaryMaterials
\section{Supplementary figures}
\begin{figure}[ht]
\center{\includegraphics[width=\textwidth]{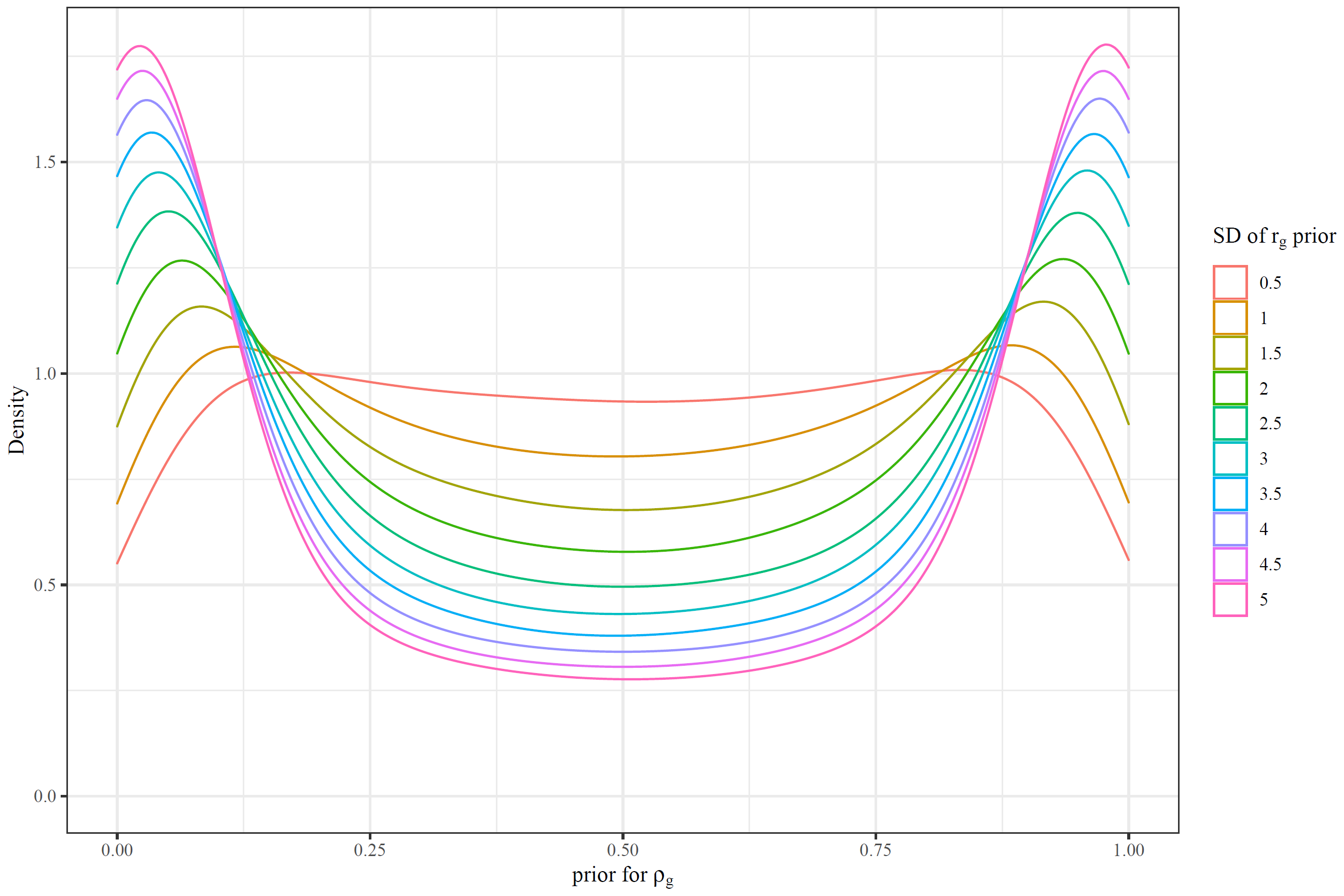}}
\caption{Marginal prior distributions of $\rho_g$ with increasing standard deviations of the normally distributed prior for $r_g$  (whose mean is 0), and a prior distribution of $U(0,1)$ for $P$. The marginal prior distribution for $\rho_g$ is approximately $U(0,1)$ when a $N(0,0.25)$ is imposed on $r_g$. Thus, constraining the prior for $r_g$ to a relatively narrow distribution  results in a diffuse marginal prior for $\rho_g$.}
\label{fig:r}
\end{figure}

\begin{figure}
\center{\includegraphics[width=\textwidth]{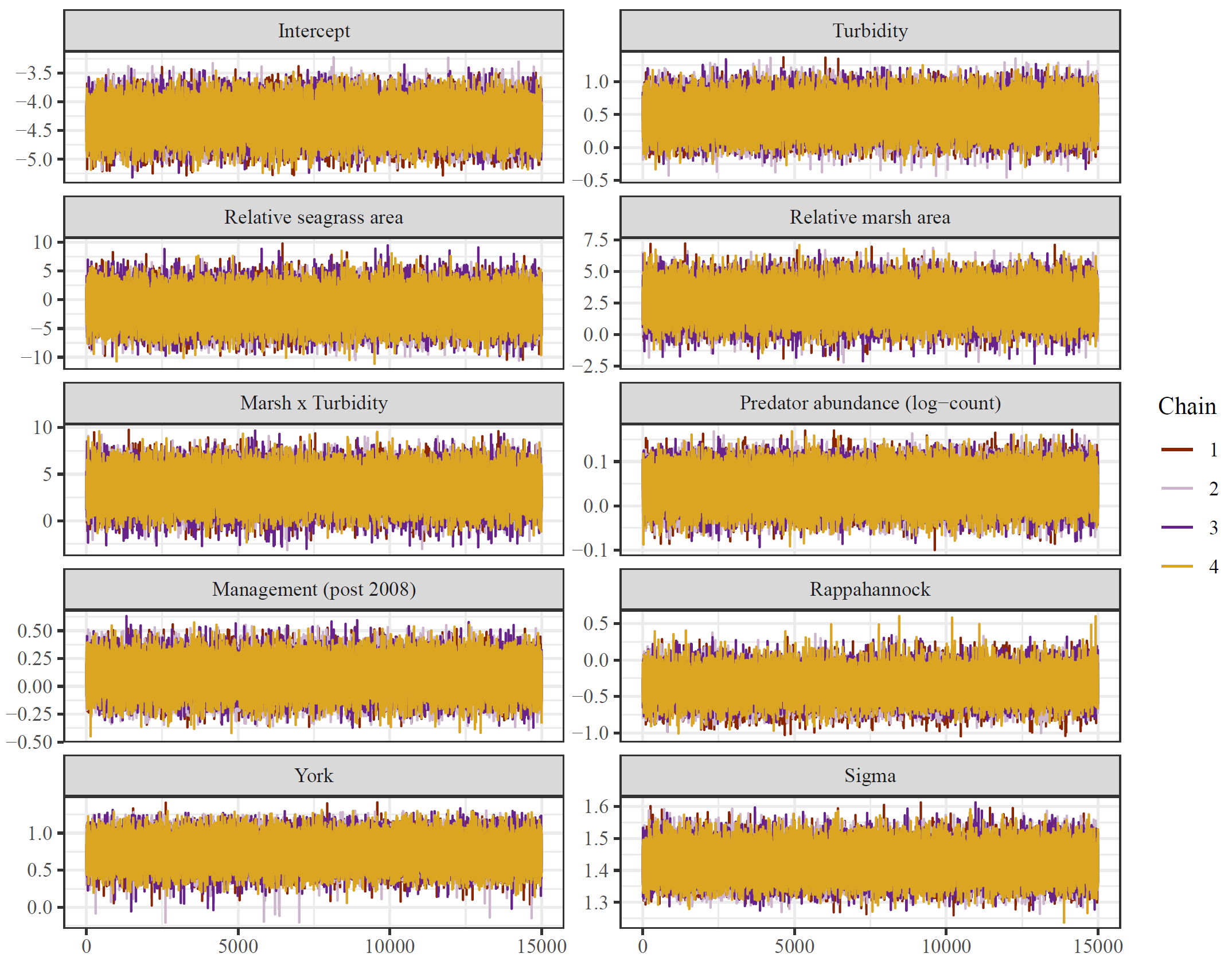}}
\caption{A set of trace plots for Model 4 parameters illustrating sampled values of each regression coefficient and $\sigma_\Phi$ per chain throughout the post burn-in iterations. Visual inspection of trace plots is used to evaluate convergence and mixing of the chains.}
\label{fig:Traceplot}
\end{figure}

\begin{figure}
\center{\includegraphics[width=\textwidth]{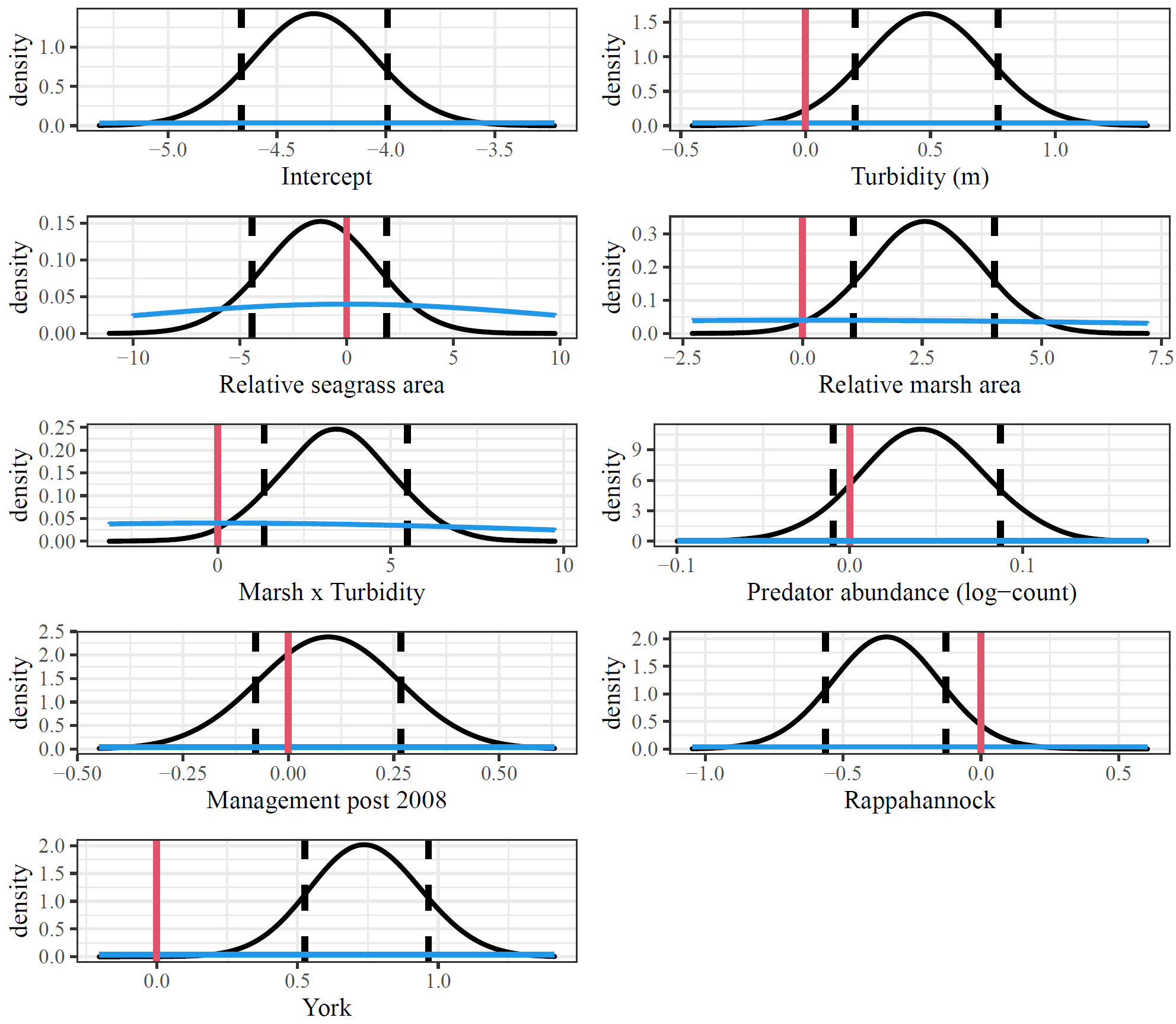}}
\caption{Posterior distributions (black) and prior distributions (blue) of regression coefficients from Model 4; dashed black lines denote 80\% credible intervals, while solid red lines denote 0}
\label{fig:Posterior1}
\end{figure}

\begin{figure}
\center{\includegraphics[width=\textwidth]{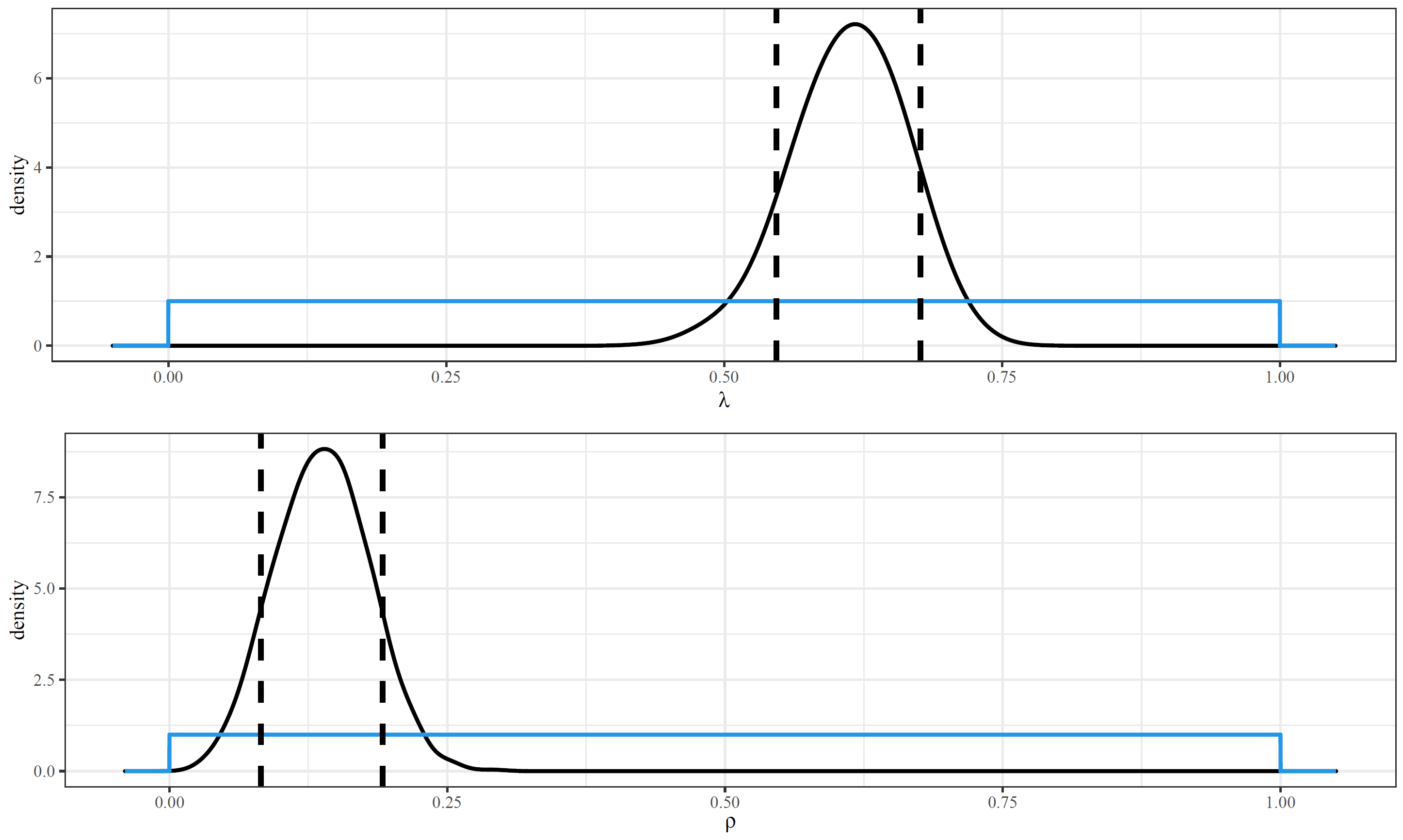}}
\caption{Posterior distributions (black) and prior distributions (blue) of autocorrelation  parameters $\lambda$ (spatial) and $\rho$ (temporal) from Model 4; dashed black lines denote 80\% credible intervals. Leave-future-out cross validation of Models 1--4 showed that the non-separable spatiotemporal dependence structure of Model 4 was necessary for good predictive performance, despite the small $\rho$.}
\label{fig:Posterior2}
\end{figure}

\begin{figure}
\center{\includegraphics[width=\textwidth]{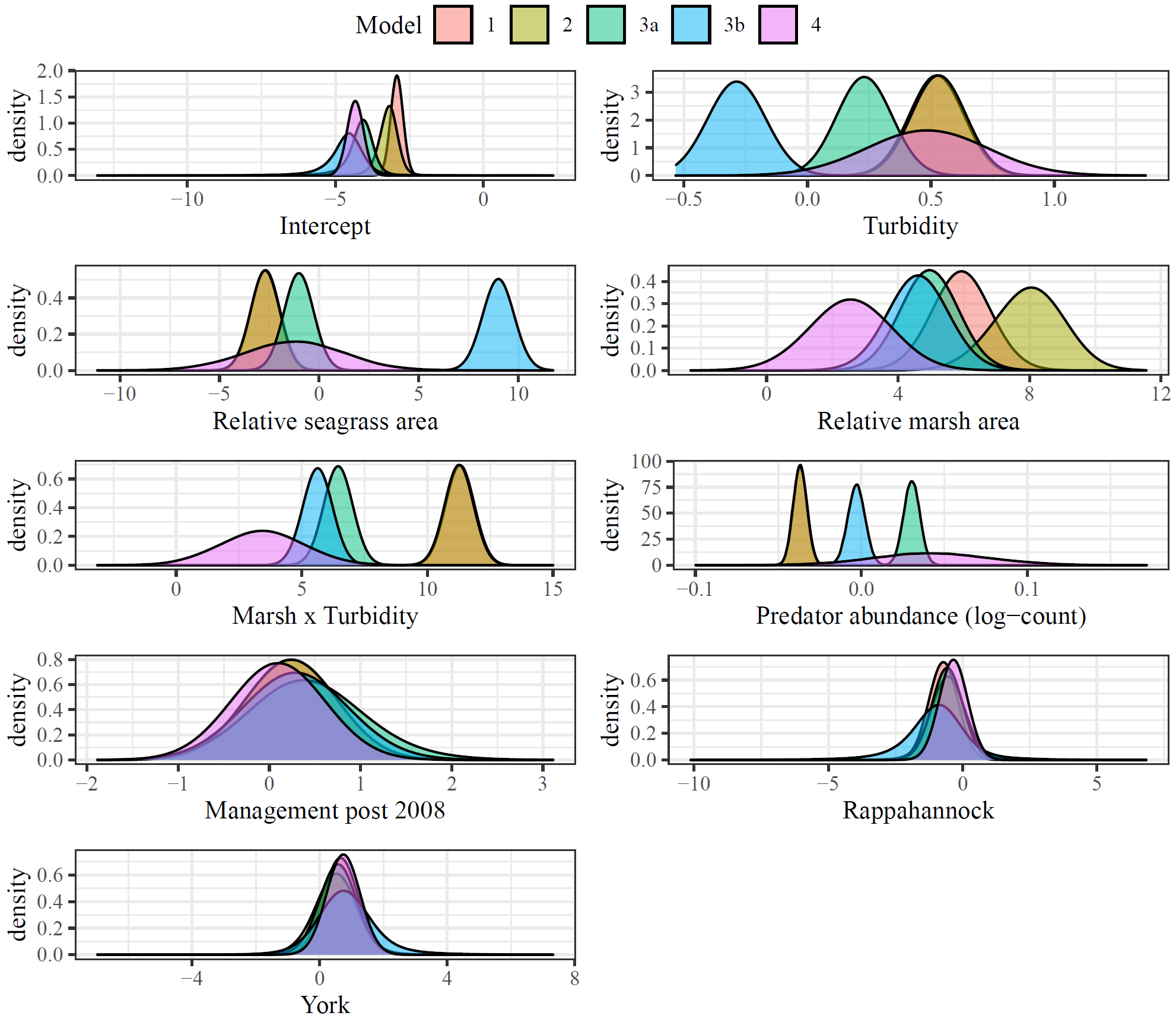}}
\caption{Posterior distributions of regression coefficients from Models 1-4.}
\label{fig:Posterior3}
\end{figure}

\end{appendix}

\clearpage

\begin{acks}[Acknowledgments]
The authors acknowledge William \& Mary Research Computing for providing computational resources and technical support that have contributed to the results reported within this paper. URL: \url{https://www.wm.edu/it/rc}. ACH also thanks D Eggleston and C Patrick for their ideas as members of ACH's PhD Committee. This project would not have been possible without the help of J Buchanan, A Comer, W. Lowery, K. Nickerson, D. Royster, and the many other individuals who collected blue crabs between 1996 and 2017 from the VIMS Juvenile Fish Trawl Survey; in particular, we thank W Lowery and T Tuckey for providing access to survey records.  

\end{acks}

\begin{funding}
Preparation of this manuscript by ACH was funded by a Willard A. Van Engel Fellowship of the Virginia Institute of Marine Science, William \& Mary.
\end{funding}


\bibliographystyle{imsart-nameyear} 
\bibliography{References}       

\end{changemargin}
\end{document}